\documentclass[usenatbib, twocolumn]{mn2e}

\usepackage{graphicx}
\usepackage{color}
\usepackage{multirow}
\usepackage{amssymb}
\usepackage{amsmath}
\usepackage{anysize}
\usepackage{aas_macros}

\renewcommand{\vec}[1]{\bmath{#1}}

\newcommand{\av}[1]{\left<#1\right>}

\title[Instabilities in rotating jets]
      {Linear stability analysis of magnetized  relativistic rotating jets}

\author[G.Bodo et al.] {G. Bodo$^{1}$\thanks{E-mail:
bodo@oato.inaf.it}, G. Mamatsashvili$^{2,3,4,5}$, P. Rossi$^{1}$ and A. Mignone$^{6}$\\
$^{1}$INAF/Osservatorio Astrofisico di Torino, Strada Osservatorio 20, 10025 Pino Torinese, Italy\\
$^{2}$Niels Bohr International Academy, Niels Bohr Institute,
Blegdamsvej 17, 2100 Copenhagen, Denmark\\
$^{3}$Helmholtz-Zentrum Dresden-Rossendorf, D-01328 Dresden, Germany\\
$^{4}$Abastumani Astrophysical Observatory, Ilia State University, G. Tsereteli str. 3, Tbilisi 0162, Georgia\\
$^{5}$Institute of
Geophysics, Tbilisi State University, Aleksidze str. 1, Tbilisi 0193, Georgia\\
$^{6}$Dipartimento di Fisica,  Universit\`a degli Studi di Torino, Via Pietro Giuria 1, 10125 Torino, Italy}

\begin{document}

\date{Accepted ??. Received ??; in original form ??}

\pagerange{\pageref{firstpage}--\pageref{lastpage}} \pubyear{2016}

\maketitle

\label{firstpage}

\begin{abstract}
We carry out a linear stability analysis of a magnetized relativistic
rotating cylindrical jet flow using the approximation of zero thermal
pressure. We identify several modes of instability in the jet:
Kelvin-Helmholtz, current driven and two kinds of
centrifugal-buoyancy modes -- toroidal and poloidal. The Kelvin-Helmholtz 
mode is found at low magnetization and its growth rate depends very weakly
on the pitch parameter of the background magnetic field and on
rotation. The current driven mode is found at high magnetization,
the values of its growth rate and the wavenumber, corresponding to
the maximum growth, increase as we decrease the pitch parameter of
the background magnetic field. This mode is stabilized by rotation, 
especially, at high magnetization. The centrifugal-buoyancy modes,
arising due to rotation, tend also to be more stable when
magnetization is increased. Overall, relativistic jet flows appear 
to be more stable with respect to their non-relativistic counterpart.
\end{abstract}


\section{Introduction}
%
%
%

The study of jet instabilities is of utmost importance for understanding their dynamics and 
phenomenology. Astrophysical jets propagate over very large distances (up to $10^9$ times 
their initial radius in the case of AGN jets) maintaining a coherent structure and, for this 
remarkable stability property, an acceptable explanation is still missing. On the other hand,  
instabilities can play a fundamental role in the dissipation of part of the jet energy, leading to the 
observed radiation as well as the formation and evolution of various observed structures. One of 
the mechanisms through which dissipation of the jet energy may occur, and that has recently 
attracted a lot of interest, is magnetic reconnection \citep[see e.g.][]{Giannios10, Sironi15, 
Werner18}. In this context, current driven kink instabilities (CDI) may play an important role  by 
enhancing or killing reconnection \citep{Striani16, Ripperda17a, Ripperda17b}. Apart from CDI, 
other types of instabilities are possible in jets: Kelvin-Helmholtz instabilities (KHI) driven by the 
velocity shear and centrifugal-buoyancy instabilities driven by the jet rotation. While the 
Newtonian, or non-relativistic case has been extensively studied, general analyses in the relativistic 
regime, without invoking the force-free approximation, i.e., taking into account gas inertia, are 
more limited due to the complexity of the problem.  By ``relativistic'' we 
mean that the Lorentz factor of the jet flow is larger than unity and/or the magnetization (i.e., the 
ratio of the magnetic energy density to the energy density of matter) is high, enabling the jet 
to accelerate to relativistic velocities.   KHI have been extensively studied in several different 
configurations both in the non-relativistic \citep[see e.g.][]{Bodo89, Birkinshaw91, Hardee92, 
Bodo96, Hardee06, Kim15} and relativistic \citep[see e.g.][]{Ferrari78, Hardee79, Urpin02, 
Perucho04, Perucho10, Hardee07} cases. Similarly, CDI have been extensively studied in the 
Newtonian limit both in the linear \citep[e.g.][]{Appl92, Appl96,Begelman98, Appl00, Baty02, 
Bonanno11, Bonanno11a, Das18} and nonlinear \citep[e.g.,][]{Moll_etal08,ONeill_etal12} regimes, 
while the analysis of the relativistic case has been more limited, most of the linear studies have 
considered the force-free regime \citep{Voslamber_Callebaut62, Pariev94, Pariev96, Lyubarski99, 
Tomimatsu01, Narayan09, Gourgouliatos_etal12, Sobacchi_etal17} and the full MHD case has been 
addressed more recently by \citet{Bodo13} (hereinafter Paper I), for the cold case, and by 
\citet{Begelman98, Kim17, Kim18}, who included thermal pressure. Due to the complexity of the 
relativistic case, the evolution of CDI beyond the force-free approximation has been tackled often 
by means of numerical simulations, which mainly focus on the nonlinear behaviour 
\citep[e.g.,][]{Mizuno_etal09, Mizuno_etal11, ONeill_etal12,Mizuno_etal12, Mignone_etal10, 
Mignone_etal13, Singh_etal16}. In the absence of magnetic fields, rotation can drive the centrifugal 
instability in jets, whose relativistic extension has been recently analysed by \citet{Komissarov18}. 
The combination of rotation and magnetic field adds another degree of complexity, other kinds of 
instabilities may arise and the interplay between the different modes can become quite 
complicated \citep[for non-relativistic studies, see e.g.][]{Kim00, Hanasz00, Keppens02, 
Varniere02, Huang03,Pessah05, Bonanno06, Bonanno07, Fu11}. The interplay of rotation and 
magnetic field, in the non-relativistic case and in the absence of a longitudinal flow, has been 
analyzed by \citet{Bodo16} (hereinafter Paper II), and the resulting main, rotationally-induced types 
of instability -- the centrifugal-buoyancy modes -- have been identified and described.

 In Paper I we considered a cold, relativistic, non-rotating jet and found that KHI is prevalent for 
matter-dominated jets, while CDI is more effective for magnetically-dominated jets. In Paper II, we 
considered the effects of rotation in a non-relativistic plasma column, where no longitudinal flow is 
present. We found additional modes of instability driven by rotation: the centrifugal-buoyancy 
modes. In this paper, which represents a sequel of Papers I and II, we study the stability problem in 
the full case of a cold, relativistic, magnetized and rotating jet. We still consider a cold jet, because, 
on one side, in the case of a Poynting-dominated jet, this can be 
assumed as a valid approximation and, on the other side, the incorporation of pressure would 
introduce new kinds of instabilities even more  complicating the analysis. This is therefore a further 
step towards a complete study, where we will drop this limitation in the end. The equilibrium 
configuration here is similar to that adopted in these papers, which assumes a current distribution 
that is peaked on the jet axis and closes at very large distances from the jet (i.e., the total net 
current becomes equal to zero only at large distances).  This class of equilibria is different from 
those considered by \citet{Kim17, Kim18}, where the current closes inside the jet. Resulting main 
modes of the instability in the relativistic and rotating case remain the KH, CD and 
centrifugal-buoyancy ones. The main goal of the present paper is to investigate the effect of the 
different parameters of the jet on the growth efficiency of these modes in the more comprehensive 
relativistic rotating case compared to the relativistic non-rotating and non-relativistic rotating ones 
analyzed, respectively, in Papers I and II.  The main parameters, with respect to which we 
explore the jet stability, are the Lorentz factor of the propagation velocity along jet axis, pitch of 
the background magnetic field, degree of magnetization, rotation frequency, vertical/axial 
wavenumber. In contrast to the present more general study, in Paper I rotation was zero, whereas 
in Paper II, being in the Newtonian limit, the Lorentz factor was unity and the magnetization, as 
defined here, was very small. Ultimately, one would like to understand how instabilities can tap part 
of the jet flow energy, without leading to its disruption. To this aim, numerical simulations are an 
essential tool, however, linear studies such as the present one may still provide necessary insights.

The plan of the paper is the following: in section 2 we will describe the physical problem, the basic 
equations, the general equilibrium configuration and the characteristic parameters, in 
section 3 we present our results, first for the KHI and CDI and then for the centrifugal-buoyancy 
instabilities and, finally in section 4, we summarize our findings.

\section{Problem Description}
\label{problem}
%
%
%
 We investigate the linear stability of a cold (i.e., with zero thermal pressure),
magnetized, rotating, relativistic cylindrical flow of an inviscid and infinitely conducting fluid. 
It is governed by the basic equations of ideal relativistic MHD:
\begin{equation}\label{eq:drho/dt}
\frac{\partial}{\partial t} (\gamma \rho) + \nabla \cdot (\gamma
\rho  \vec{v}) = 0,
\end{equation}
\begin{equation}\label{eq:dm/dt}
\gamma \rho \frac{\partial}{\partial t}(\gamma \vec{v} ) + \gamma
\rho (\vec{v} \cdot \nabla ) (\gamma \vec{v} ) = \frac{1}{c}\vec{J} \times
\vec{B} + \frac{1}{4\pi}( \nabla \cdot \vec{E} ) \vec{E},
\end{equation}
\begin{equation}\label{eq:dB/dt}
\frac{1}{c}\frac{\partial\vec{B}}{\partial t} = - \nabla \times \vec{E},
\end{equation}
\begin{equation}\label{eq:dE/dt}
\frac{1}{c}\frac{\partial \vec{E}}{\partial t} = \nabla \times \vec{B} - \frac{4\pi}{c}\vec{J},
\end{equation}
where $\rho$ is the proper density, $\gamma=(1-v^2/c^2)^{-1/2}$ is
the Lorentz factor, with $c$ being the speed of light, and $\vec
{v}$, $\vec {B}$, $\vec{E}$, $\vec{J}$ are, respectively, the 3-vectors of the
velocity, magnetic field, electric field and current density. These equations 
are written in the CGS system and a factor of $\sqrt{4 \pi}$ is
absorbed in the definitions of $\vec{E}$ and $\vec{B}$. In the following we choose the units  
such that the speed of light is unity, $c = 1$, .

\subsection{Equilibrium Configuration}
%
%
\label{sec:equilibrium}

The equilibrium configuration was described in Paper I, here we
summarize the relevant equations. We adopt cylindrical coordinates
$(r,\varphi,z)$ (with versors $\vec{e}_r, \vec{e}_\varphi,
\vec{e}_z$) and seek for axisymmetric steady-state solutions (i.e.,
$\partial_t=\partial_\varphi=\partial_z=0$) of equations
(\ref{eq:drho/dt})-(\ref{eq:dE/dt}). The jet propagates in the
vertical/axial ($z$) direction, the magnetic field and velocity have no
radial components and consist of a vertical (poloidal), $B_z, v_z$,
and toroidal, $B_\varphi, v_\varphi$, components. The magnetic field
configuration can be characterized by the pitch parameter
\[
P=\frac{rB_z}{B_{\varphi}}.
\]
The only non-trivial equation is given by the radial component of
the momentum equation (\ref{eq:dm/dt}) which, in the zero pressure
case, simplifies to
\begin{equation}\label{eq:radial_eq}
\rho\gamma^2v_\varphi^2 =
\frac{1}{2r}\frac{d(r^2H^2)}{dr}+\frac{r}{2}\frac{dB_z^2}{dr},
\end{equation}
where $H^2 = B_\varphi^2 - E_r^2$ (in the non-relativistic case $H^2 = B_\varphi^2$) and
\begin{equation}\label{eq:elec_field}
E_r = v_z B_\varphi-v_\varphi B_z
\end{equation}
Equation (\ref{eq:radial_eq}) leaves the freedom of choosing the
radial profiles of all flow variables except one and then solve for
the remaining profile. We note that while in the Newtonian case, the
presence of a longitudinal velocity has no effect on the radial
equilibrium, this no longer holds in the relativistic case, where
the Lorentz factor appears in the equilibrium condition
(\ref{eq:radial_eq}). The choice of the radial profiles is somewhat
arbitrary since we have no direct information about the magnetic
configuration in astrophysical jets. We choose to follow the
prescriptions given in Papers I and II and to consider a general
class of constant density equilibria in which the vertical current
density is peaked on the central axis of the jet and is concentrated in a
region of the characteristic radius $a$. We prescribe the velocity profile by choosing
$\gamma_z(r)$, i.e., the Lorentz factor with respect to the $z$-component of the velocity only, of the form
\begin{equation}\label{eq:vz_prof}
 \gamma_z(r) \equiv \frac{1}{\sqrt{1-v_z^2}}  =  1 + \frac{\gamma_c - 1}{\cosh(r/r_j)^6},
\end{equation}
where $\gamma_c = (1-v_c^2)^{-1/2}$ is the Lorentz factor for the vertical velocity on the 
central axis, $v_c=v_z(0)$, and $r_j$ is the jet radius. From now on, we will use the
subscript $`c`$ to denote values at $r=0$, in addition, all lengths
will be expressed in units of $r_j$ (recall that the velocities are
measured in units of the speed of light $c$). As in Paper I, we
prescribe the profile of $H$ as
\begin{equation} \label{eq:H2_prof}
 H^2 = \frac{H^2_c}{r^2}\left[1 - \exp\left(-\frac{r^4}{a^4}\right)\right]
\end{equation}
and for the azimuthal velocity we take the form
\begin{equation} \label{eq:vphi_prof}
\gamma^2  v_\varphi^2 =  r^2 \Omega^2_c \gamma_c^2
\exp\left(-\frac{r^4}{a^4}\right),
\end{equation}
where  $\Omega_c$ is the angular velocity of the jet rotation on the central axis 
($\gamma_c\Omega_c$ is the angular velocity measured in the jet rest frame). 
 The characteristic radius of the current concentration in the jet is set to $a=0.6$ below.
With these choices, from equation (\ref{eq:vphi_prof}) we get for $v_\phi$ the expression
\begin{equation}\label{eq:vphi2_prof}
v_\varphi^2 = \frac{r^2\gamma_c^2\Omega_c^2}{\gamma_z^2} \left[1
+r^2 \gamma_c^2 \Omega_c^2 \exp \left( -\frac{r^4}{a^4}\right)
\right]^{-1}\exp \left(-\frac{r^4}{a^4}\right),
\end{equation}
from which it is evident that for any value of $\Omega_c$, $v_\phi$
is always less than unity, i.e., the azimuthal velocity does not
exceed the speed of light. From equations (\ref{eq:radial_eq}),
(\ref{eq:H2_prof}) and (\ref{eq:vphi_prof}), we get the $B_z$
profile as
\begin{equation}\label{eq:Bz_prof}
B^2_z = B^2_{zc} - (1 - \alpha) \frac{H^2_c\sqrt{\pi}}{a^2}{\rm erf}
\left(\frac{r^2}{a^2}\right)
\end{equation}
where $\mathrm{erf}$ is the error function and the parameter
\begin{equation}\label{eq:alfa}
\alpha = \frac{\rho\gamma_c^2\Omega^2_ca^4}{2H_c^2}
\end{equation}
measures the strength of rotation: for $\alpha = 0$ (no rotation)
the gradient of $r^2 H^2$ in equation (\ref{eq:radial_eq}) is
exactly balanced  by the gradient of $B_z^2$ ($B_z$ decreases
outward), whereas for $\alpha = 1$, it is exactly balanced by the
centrifugal force and $B_z$ is constant. Intermediate values of
rotation correspond to  the range $0<\alpha<1$. As shown in Paper
II, one can, in principle, consider also configurations with $\alpha
> 1$, in which $B_z$ grows radially outward, but such configurations
will not be considered in the present paper.

The azimuthal field is obtained from the definition of $H$ using the
expression of $E_r=v_zB_\varphi-v_\varphi B_z$. This yields a
quadratic equation in $B_\varphi$ with the solution
\begin{equation}\label{eq:Bphi}
B_\varphi = \frac{-v_\varphi v_zB_z\mp\sqrt{v_\varphi^2B_z^2 +
H^2(1-v_z^2)}}{1-v_z^2}.
\end{equation}
Here we consider the negative branch because it guarantees that
$B_\varphi$ and $v_\varphi$ have opposite signs, as suggested by
acceleration models  \citep[see e.g.,][]{Blandford_Payne82, Ferreira_Pelletier95, 
Zanni_etal07}. We choose to characterize the magnetic field configuration by specifying the 
absolute value of the pitch on the axis, $P_c$, and the ratio of the energy density of the 
matter to the magnetic energy density, $M_a^2$,
\begin{equation}\label{eq:P_cnd_MA}
P_c \equiv \left|\frac{rB_z}{B_\varphi}\right|_{r=0} \,,\qquad M_a^2
\equiv \frac{\rho \gamma_c^2}{\av{\vec{B}^2}} \,,
\end{equation}
where $\av{\vec{B}^2}$ represents the radially averaged magnetic energy density across the beam
\begin{equation}\label{eq:Bav}
\av{\vec{B}^2} = \frac{\int_0^{r_j} (B_z^2 + B_\varphi^2)r\,dr}
{\int_0^{r_j} r \,dr},
\end{equation}
and $r_j=1$ in our units.  $M_a$ is related to the standard magnetization 
parameter $\sigma=B^2/(\rho h)$ ($h$ is the specific enthalpy) 
used in other studies via $M_a^2=\gamma_c^2/\sigma$ and 
to the relativistic form of the Alfv\'en speed, $v_a=B/\sqrt{\rho h+B^2}$, 
via $M_a^2=\gamma_c^2(1-v_a^2)/v_a^2$ ($c=1$ and in the cold limit $h=1$). 
The constants $B_{zc}$ and $H_c$ appearing in 
the above equations can be found in terms of 
$P_c$, $M_a$ and $\Omega_c$ by simultaneously solving equations
(\ref{eq:P_cnd_MA}) and (\ref{eq:Bav}) using expressions
(\ref{eq:Bz_prof}) and (\ref{eq:Bphi}) for the magnetic field
components. In particular, from the definition of the pitch
parameter, after some algebra, we find (in the $r\rightarrow 0$
limit)
\begin{equation}
a^4 B^2_{zc} = \frac{H_c^2 P_c^2}{1-  \left( P_c\Omega_c + v_{zc}
\right)^2}.
\end{equation}
\begin{figure*}
\centering
\includegraphics[width=12cm]{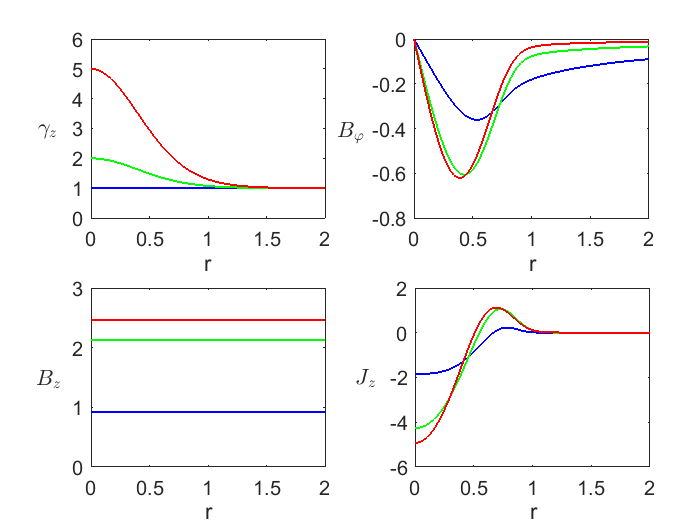}
\caption{Radial profiles of the Lorentz factor, magnetic field components and axial current 
density for the equilibrium at $\alpha=1$, $M_a^2=1$, $P_c=1$ and different $\gamma_c=1.01 
~(blue), 2~(green), 5~(red)$.}\label{fig:radial}
\end{figure*}
Fig. \ref{fig:radial} shows the typical radial profile of this equilibrium solution for the 
special/representative case of maximal rotation, $\alpha=1$, and different $\gamma_c$. However, 
as discussed in Paper II for the non-relativistic case, not
all the combinations of $\Omega_c$, $\gamma_c$, $P_c$ and $M_a$ are
allowed, because, in order to have a physically meaningful solution,
we have to impose the additional constraints that $B_\varphi^2$ and
$B_z^2$ must be everywhere positive. We note that, since $B_z^2$
decreases with radius monotonically, the condition $\lim_{r
\rightarrow \infty}B_z^2 > 0$ ensures that $B_z^2$ is positive
everywhere.
\begin{figure*}
\centering
\includegraphics[width=5cm]{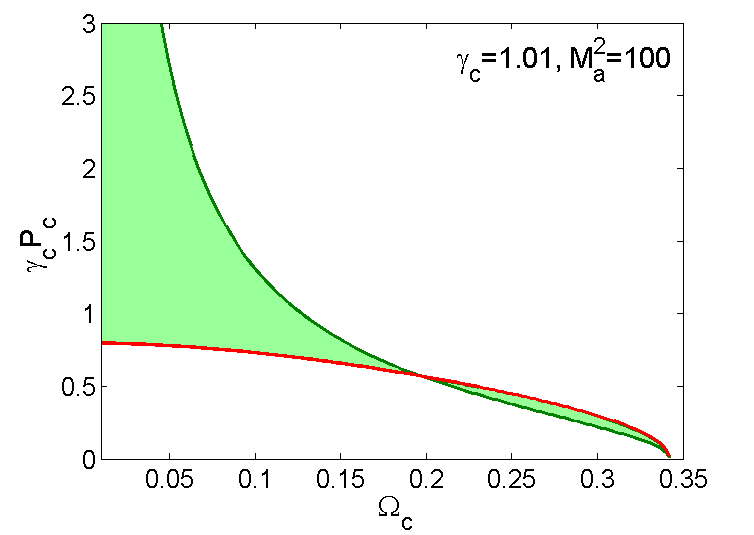} 
\includegraphics[width=5cm]{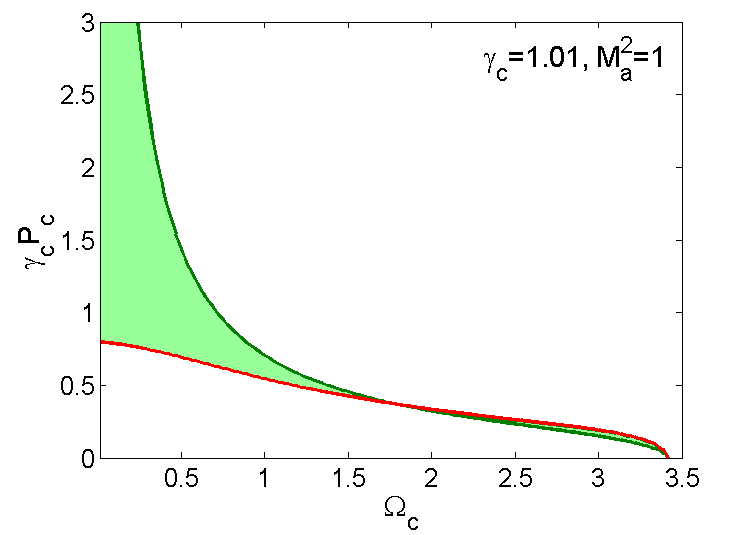} 
\includegraphics[width=5cm]{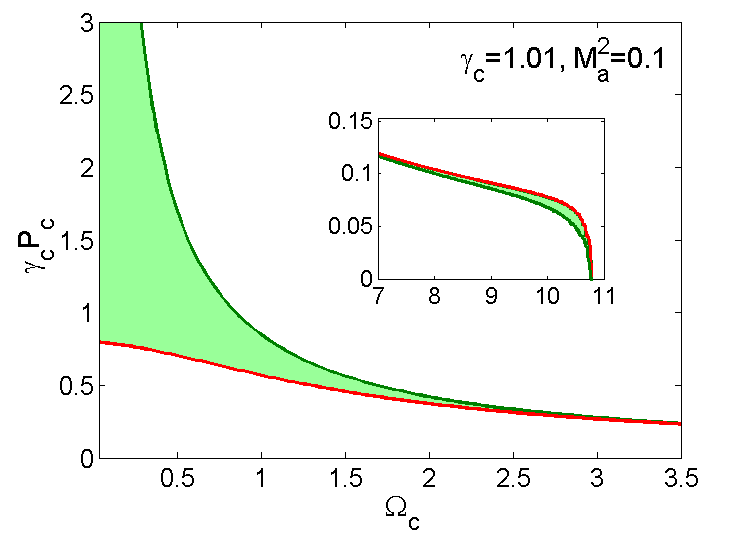} 
\includegraphics[width=5cm]{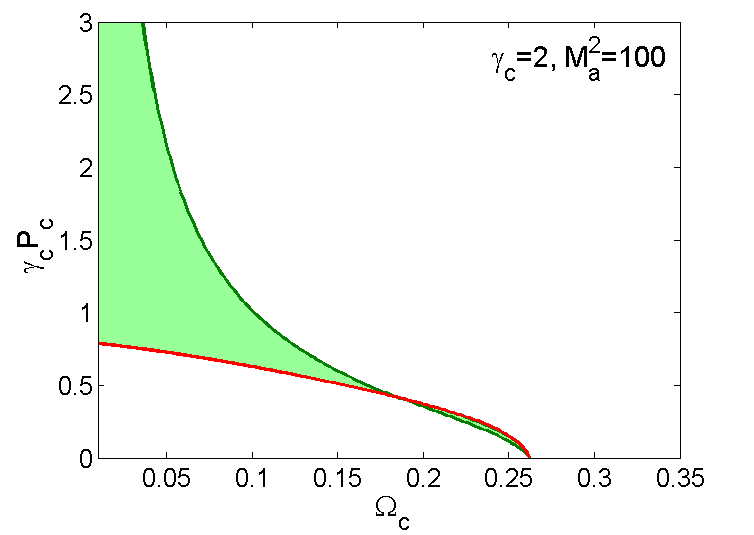} 
\includegraphics[width=5cm]{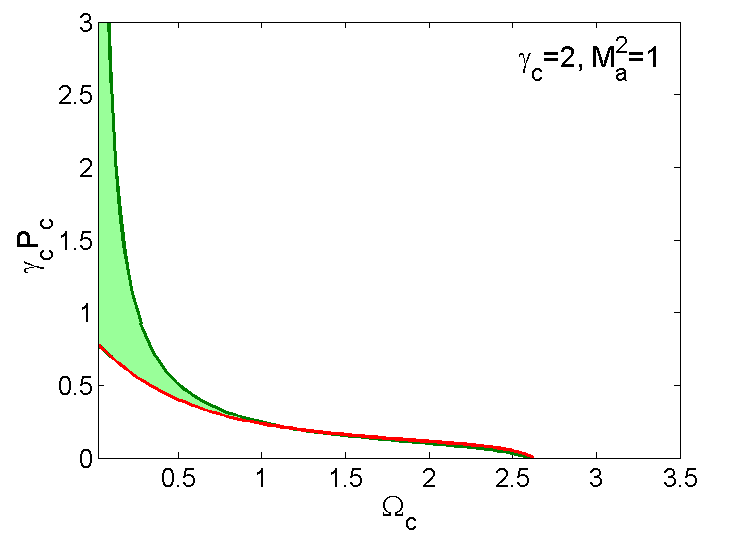} 
\includegraphics[width=5cm]{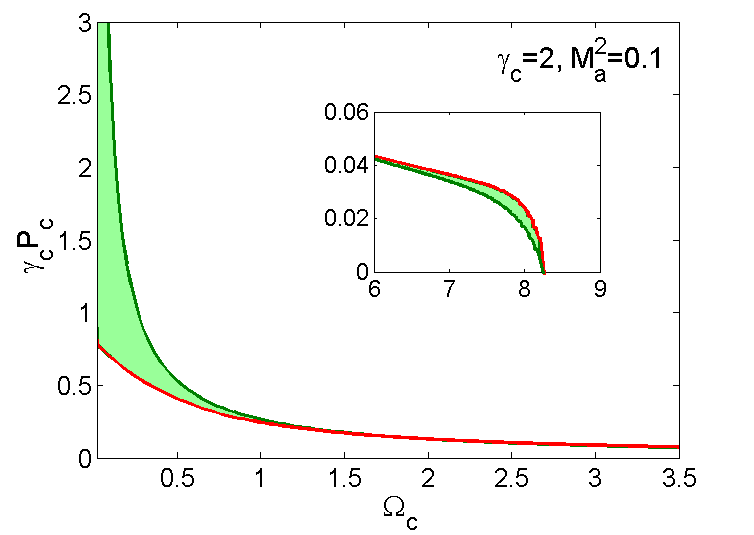} 
\includegraphics[width=5cm]{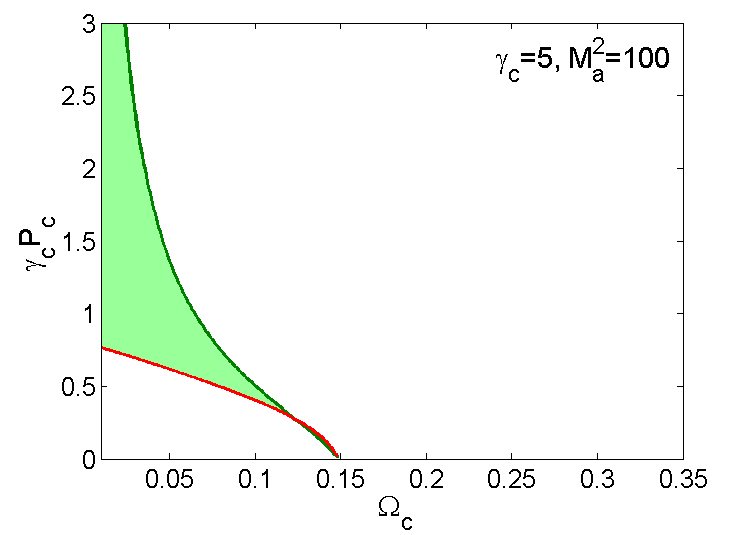} 
\includegraphics[width=5cm]{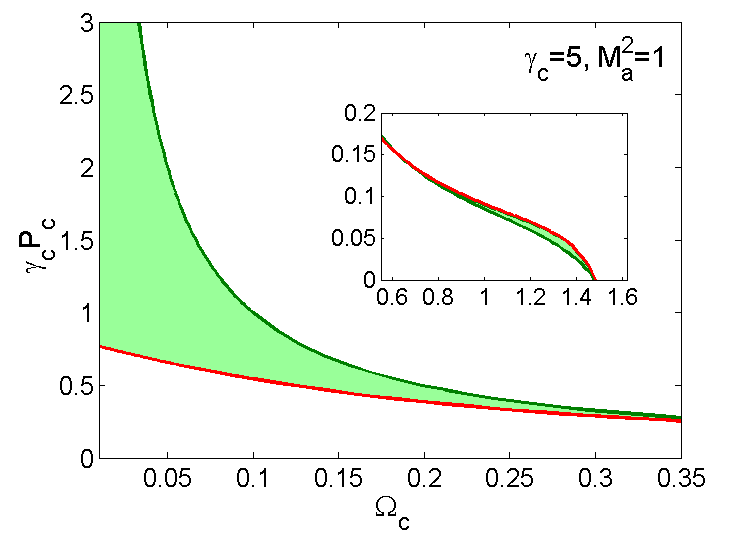} 
\includegraphics[width=5cm]{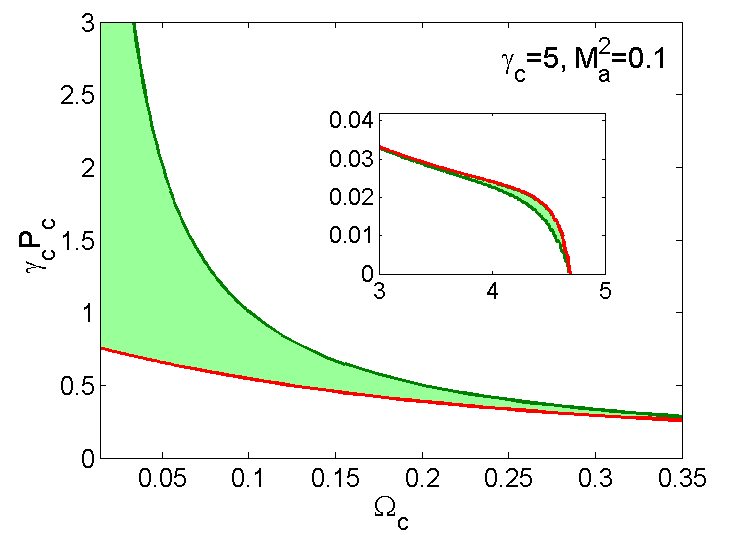} 
\caption{\small Regions of allowed equilibria in the ($\Omega_c,
\gamma_c P_c$)-plane shaded in light green. The different panels
refer to different values of $\gamma_c$ and $M_a$, the values
corresponding to each panel are reported in the legend. The red
curves mark the boundary where $\lim_{r \rightarrow \infty}B_z^2 =
0$ and the green curves mark the boundary where $B_z^2$ is constant
with radius ($\alpha = 1$).  Insets show the maximum $\Omega_c$ of the
possible equilibria when the latter extend beyond the range of $\Omega_c$ 
represented in these plots.}\label{fig:equil}
\end{figure*}
Fig. \ref{fig:equil} shows the allowed region (shaded in light
green) in the ($\Omega_c, \gamma_c P_c$)-plane, with the red curve
marking the boundary where $\lim_{r \rightarrow \infty}B_z^2 = 0$
and the green curve marking the boundary where $B_z^2$ becomes
constant with radius (i.e., $\alpha = 1$). (As discussed in Paper
II, there are also possible equilibria where $B_z^2$ increases with
radius, outside the green curve, but they are not considered here.)
We used $\gamma_c P_c$ on the ordinate axis, since it represents the
pitch measured in the jet rest frame. The three panels in each row
refer to decreasing values of $M_a^2$ (from left to right, $M_a^2 =
100, 1, 0.1$) corresponding to increasing strength of the magnetic
field, while the three panels in each column refer to increasing
value of the Lorentz factor (from top to bottom, $\gamma_c = 1.01,
2, 5$). The top leftmost panel with the lowest magnetization and $\gamma_c=1.01$
should correspond to the Newtonian limit shown in Fig. 1 of Paper II,
comparing these two figures we can see that the shape of the permitted
light green regions are nearly the same, while the values of $\Omega_c$ are different because of the different normalization, in Paper II it was normalized by $v_a/r_j$, whereas here it is normalized by $c/r_j$.  Even at this nearly Newtonian value of $\gamma_c=1.01$, the relativistic effects become noticeable starting from intermediate magnetization $M_a^2 = 1$ -- the corresponding permitted light green region with its green and red boundaries (top middle panel) differs from those in the Newtonian limit $M_a^2=100$ (top left panel), after taking into account the above normalization of the angular velocity.  Generally, in Fig. \ref{fig:equil}, relativistic effects become increasingly stronger, on one hand, going from left to right, because the Alfv\'en speed approaches the speed of light, and on the other hand, going from top to bottom, because the jet velocity approaches the speed of light.

At high values of the pitch, there is a maximum allowed value of
$\Omega_c$, while decreasing the pitch we see that below a threshold
value $\gamma_c P_c = 0.8$, the jet must rotate in order to ensure a
possible equilibrium and the rotation rate has to increase as the
pitch decreases. For low values of $P_c$, the allowed range of $\Omega_c$ therefore tends to
become very narrow. Comparing the panels in the three columns, we see that
increasing the magnetic field, the maximum $\Omega_c$, found for
$P_c \rightarrow 0$, increases, scaling with $1/M_a$, and the
rotation velocities become relativistic. Increasing the value of
$\gamma_c$ (middle and bottom panels), the allowed values of
$\Omega_c$ in the laboratory frame (shown in the figure) decrease,
however, they increase when measured in the jet rest frame.
\begin{figure*}
\centering
\includegraphics[width=12cm]{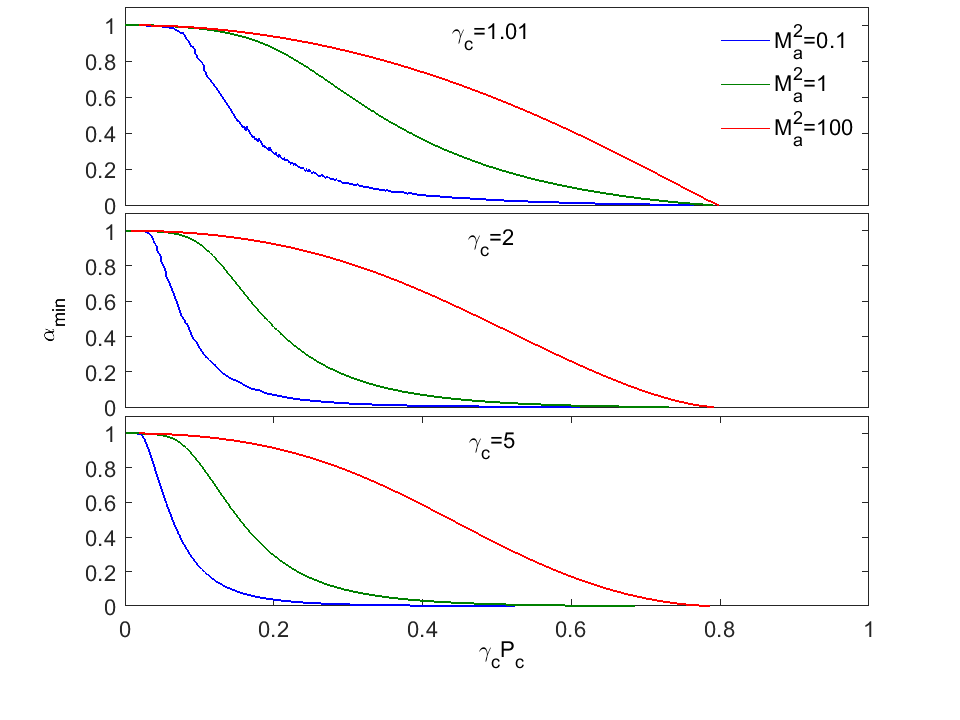} 
\caption{\small  Plots of $\alpha_{min}$ as a function of $\gamma_c
P_c$. $\alpha_{min}$ represents the minimum value of $\alpha$ for
which equilibrium is possible. The three panels refer to three
different values of $\gamma_c$, the corresponding values are
reported in each panel. The different curves correspond to different
values of $M_a$ as indicated in the legend.} \label{fig:alphamin}
\end{figure*}

Since in the stability analysis we will often make use of the
parameter $\alpha$, defined in equation (\ref{eq:alfa}), for
characterizing the equilibrium solutions, in Fig. \ref{fig:alphamin}
we show the minimum value of $\alpha$ required for the existence of
the equilibrium as a function of $\gamma_c P_c$. We recall that
$\alpha=0$ corresponds to no rotation and $\alpha=1$ corresponds to
the case where $B_z$ is constant and the hoop stresses by $B_\phi$
are completely balanced by rotation. As discussed above, for
$\gamma_c P_c < 0.8$ some rotation is needed for maintaining the
equilibrium and this minimum rotation corresponds to $\alpha_{min}$
plotted in Fig. \ref{fig:alphamin}. The three panels refer to three
different values of $\gamma_c$ and in each panel the three curves
correspond to three different values of $M_a$. Decreasing $\gamma_c
P_c$ below the critical value, $\alpha_{min}$ increases tending to 1
as $\gamma_c P_c \rightarrow 0$. Comparing the curves for the same
value of $\gamma_c$ in each panel, we see that $\alpha_{min}$
decreases as $M_a$ is decreased. Comparing the different panels, the
corresponding curves for the same value of $M_a$ also show a
decrease of $\alpha_{min}$ with $\gamma_c$.

\section{Results}
\label{results}
%
%
In Paper II, we identified and described different linear modes of
instability existing in a non-relativistic rotating static (with $v_z=0$) column:
the CD mode as well as the toroidal and poloidal buoyancy modes
driven by the centrifugal force due to rotation. In the present
analysis, we have to consider additionally the instabilities driven
by the velocity shear between the jet and the ambient medium, that
is, KH modes. We will investigate these different perturbation modes
in the following subsections.

The small perturbations of velocity and magnetic field about the
above-described equilibrium are assumed to have the form $\propto \exp({\rm
i}\omega t-{\rm i}m\varphi-{\rm i}kz)$, where the azimuthal
(integer) $m$ and axial $k$ wavenumbers are real, while
the frequency $\omega$ is generally complex, so that there is instability if
its imaginary part is negative, ${\rm Im}(\omega)<0$,  and the growth rate of the 
instability is accordingly given by $-{\rm Im}(\omega)$. The related
eigenvalue problem for $\omega$ -- the linear differential equations
(with respect to the radial coordinate) for the perturbations
together with the appropriate boundary conditions in the vicinity of
the jet axis, $r\rightarrow 0$, and far from it, $r\rightarrow
\infty$ -- were formulated in Paper I in a general form for
magnetized relativistic rotating jets, but then only the
non-rotating case was considered. For reference, in Appendix A, we
give the final set of these main equations (\ref{eq:dxi/dr}) and
(\ref{eq:dPhii/dr}) with the boundary conditions (\ref{eq:bound_in}) and (\ref{eq:bound_out}), which are solved in the present rotating case
and the reader can consult Paper I for the details of the
derivation. In this study, we focus on $m=1$ modes for the following
reasons. For CDI, this kink mode is the most effective one, leading
to a helical displacement of the whole jet body, while CDI is absent
for $m=-1$ modes (Paper I). As for KHI, it is practically
insensitive to the sign of $m$ (Paper I), so we can choose only
positive $m$. Finally, as we have seen in Paper II, the centrifugal-buoyancy 
modes also behave overall similarly at $m=-1$ and $m=1$ for large and 
small $k$, which are the main areas of these modes’ activity.

Our equilibrium configuration depends on the four  parameters $\gamma_c$, $\alpha$, $P_c$ 
and $M_a$, specifying, respectively, the jet bulk flow velocity along the axis, the strength of the centrifugal force, the magnetic pitch and the magnetization. As mentioned above, relativistic effects become important either at high values of $\gamma_c$, because the jet velocity approaches the speed of 
light, or at low values of $M_a$, because the Alfv\'en speed approaches the speed of light, even 
when $\gamma_c\sim 1$. For some of the parameters we are forced to make a choice of few 
representative values since it would be impossible to have a full coverage of the four-dimensional 
parameter space. For $\gamma_c$ we choose one value to be $1.01$ since at large $M_a$ we 
make connection with the non-relativistic results (Paper II), while at low $M_a$ we can explore the 
relativistic effects due to the high magnetization. As another value, we choose $\gamma_c = 10$, 
which can be considered as representative of AGN jets \citep{Padovani_Urry92, Giovannini_etal01, 
Marscher06, Homan12} (except for some cases in which lower values are used, since the growth 
rates of the modes for $\gamma_c = 10$ becomes extremely low). For $\alpha$ we chose the two 
limiting cases $\alpha = 0$ (no rotation) and $\alpha = 1$ (centrifugal force exactly balances 
magnetic forces) and one intermediate value, $\alpha = 0.2$, for which the effects due to rotation 
start to be substantial (notice that the relation between $\alpha$ and the rotation rate is not 
linear). Finally, for $P_c$ we explore several different values, $P_c = 0.01, \, 0.1, \, 1, \, 10$, 
depending on the allowed equilibrium configurations (see discussion above).

\begin{figure}
\centering
\includegraphics[width=\columnwidth]{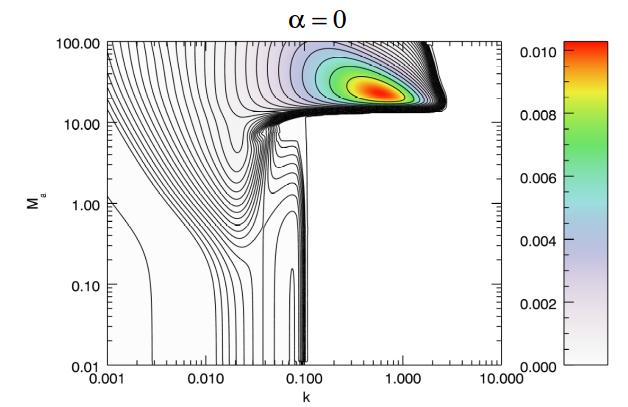} 
\includegraphics[width=\columnwidth]{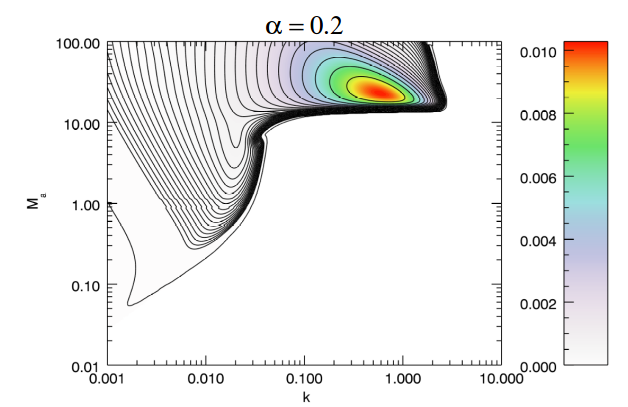} 
\includegraphics[width=\columnwidth]{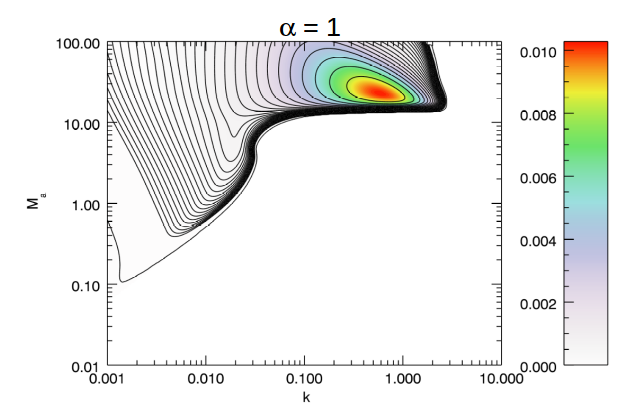} 
\caption{\small Distribution of the growth rate, $-{\rm
Im}(\omega)$, as a function of the wavenumber $k$ and $M_a$ for
$\gamma_c = 1.01$ and $P_c = 10$. The three panels correspond to
three different values of $\alpha$ (Top panel: $\alpha=0$; Middle
panel: $\alpha=0.2$; Bottom panel: $\alpha=1$).  In this and other analogous plots below, 
the colour scale covers the range from 0 to the maximum value of the growth rate, while the 
contours are equispaced in logarithmic scale from $10^{-5}$ up to this maximum growth rate.} 
\label{fig:cdvar1}
\end{figure}
\begin{figure}
\centering
\includegraphics[width=\columnwidth]{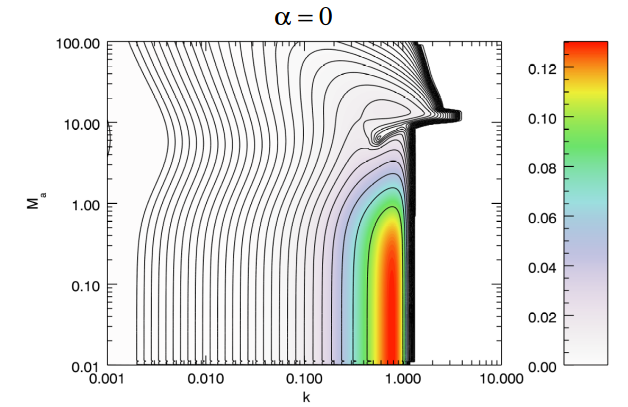} 
\includegraphics[width=\columnwidth]{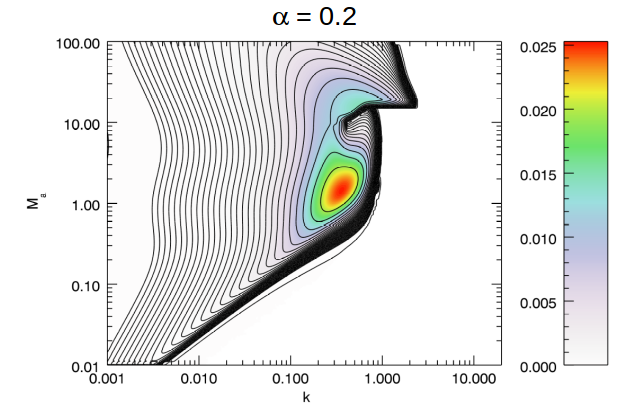} 
\includegraphics[width=\columnwidth]{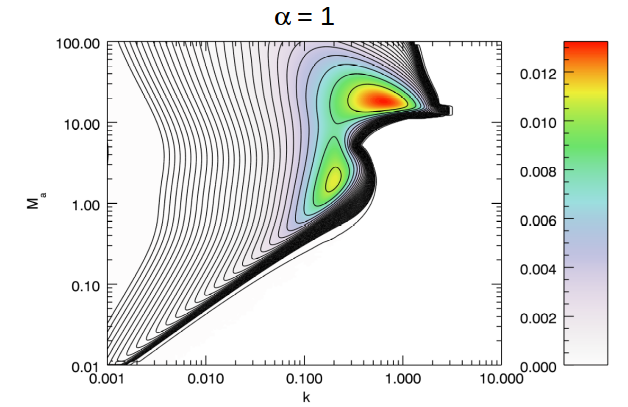} 
\
\caption{\small Same as in Fig. \ref{fig:cdvar1}, but for $\gamma_c
= 1.01$ and $P_c = 1$.}\label{fig:cdvar2}
\end{figure}
\begin{figure}
\centering
\includegraphics[width=\columnwidth]{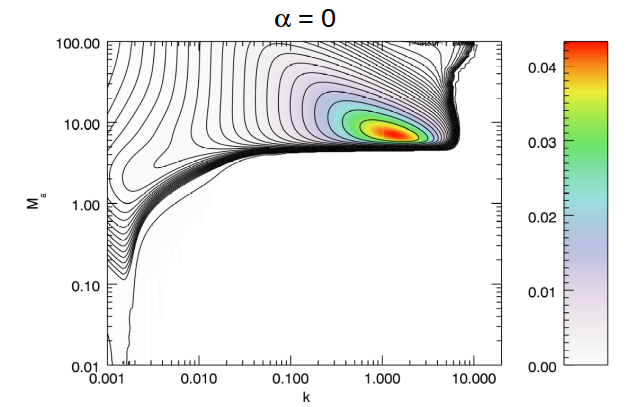} 
\includegraphics[width=\columnwidth]{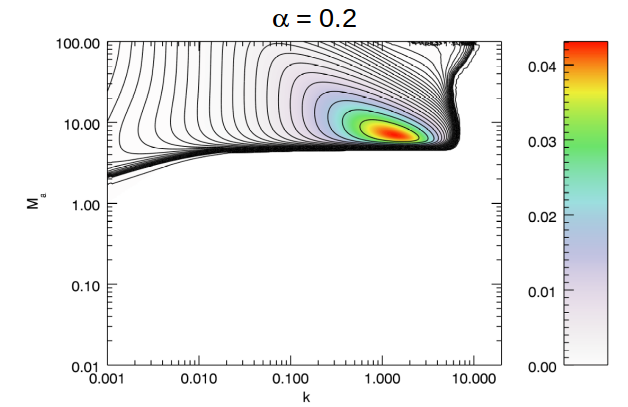} 
\includegraphics[width=\columnwidth]{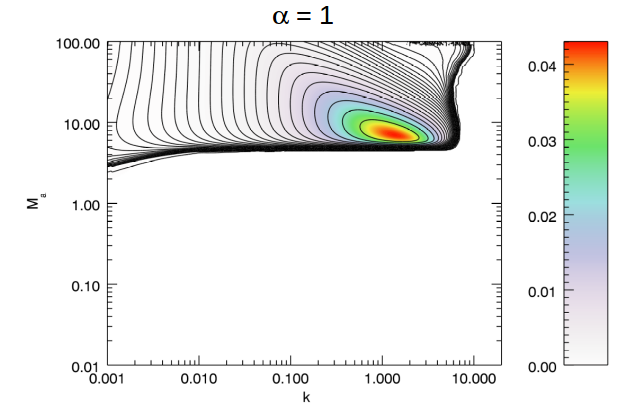} 
\
\caption{\small Same as in Fig. \ref{fig:cdvar1}, but for $\gamma_c
= 10$ and $P_c = 10$.}\label{fig:cdvar3}
\end{figure}
\begin{figure}
\centering
\includegraphics[width=\columnwidth]{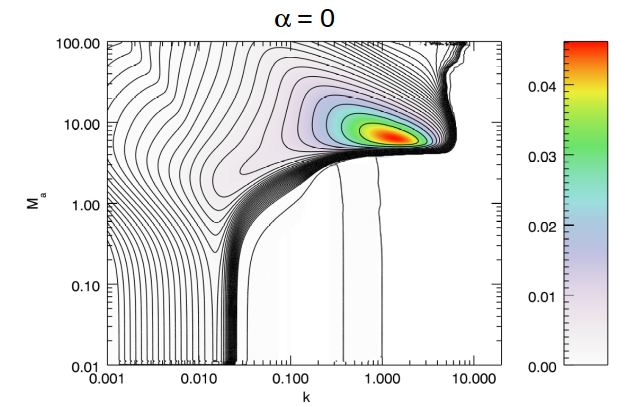} 
\includegraphics[width=\columnwidth]{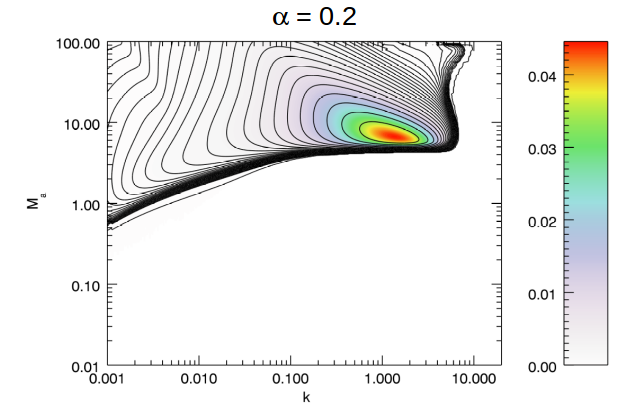} 
\includegraphics[width=\columnwidth]{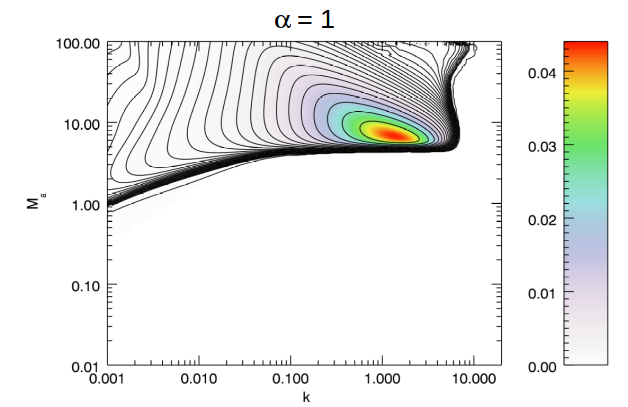} 
\caption{\small Same as in Fig. \ref{fig:cdvar1}, but for $\gamma_c
= 10$ and $P_c = 1$.} \label{fig:cdvar4}
\end{figure}
\begin{figure}
\centering
\includegraphics[width=\columnwidth]{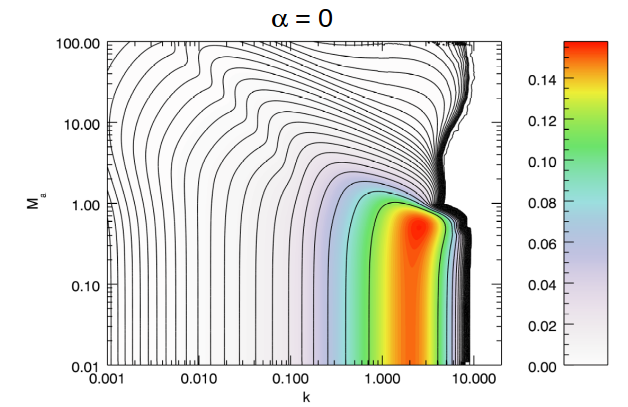} 
\includegraphics[width=\columnwidth]{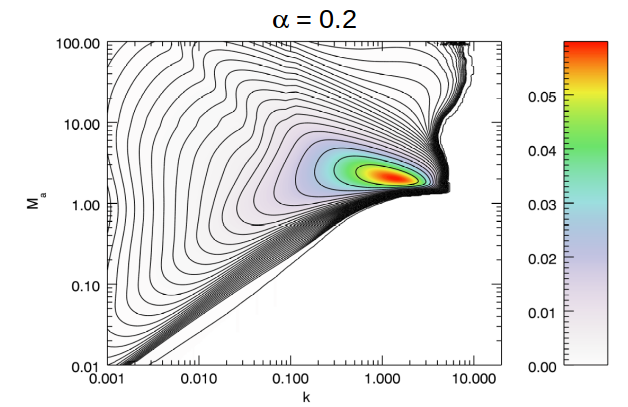} 
\includegraphics[width=\columnwidth]{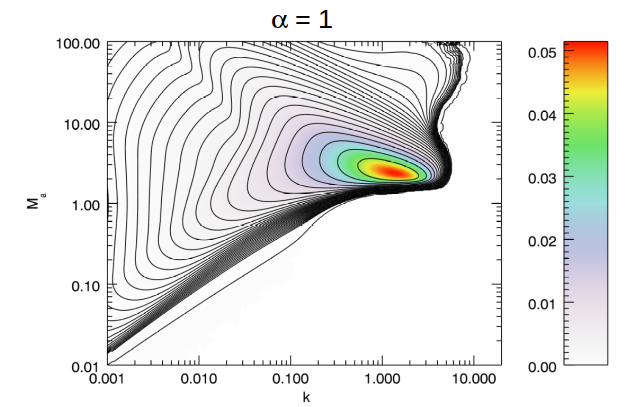} 
\caption{\small Same as in Fig. \ref{fig:cdvar1}, but for $\gamma_c
= 10$ and $P_c = 0.1$.} \label{fig:cdvar5}
\end{figure}
\begin{figure}
\centering
\includegraphics[width=\columnwidth]{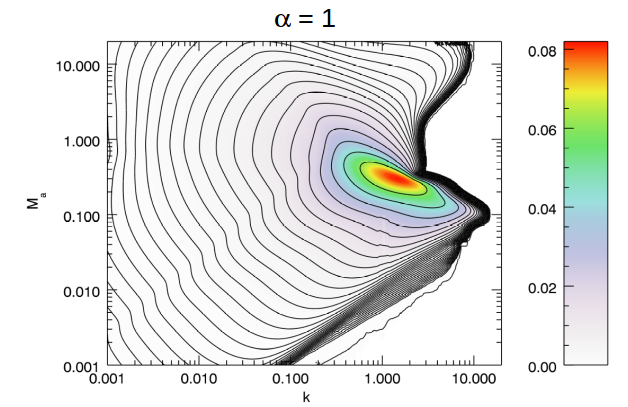} 
\caption{\small Distribution of the growth rate, $-{\rm Im}(\omega)$, as a function of the
wavenumber $k$ and $M_a$ for $\gamma_c = 10$, $\alpha = 1$ and $P_c
= 0.01$.}\label{fig:cdvar6}
\end{figure}

\subsection{CD and KH modes} 

CD and KH modes were already discussed in detail in Paper I, where
we found KH modes dominating at large values of $M_a$ and CD modes
dominating at small values of $M_a$. Here we are mainly interested in how they are affected by rotation. 
In Fig. \ref{fig:cdvar1}, we show the behaviour of the growth rate  (defined as $-{\rm Im}(\omega)$) as a function of the wavenumber $k$ and $M_a$ for $P_c=10$, $\gamma_c =1.01$ and three different values of 
$\alpha$:  $\alpha = 0$ in the top panel corresponds to no rotation, $\alpha = 1$ in the bottom  
panel corresponds to the case where rotation exactly balances magnetic forces and as an 
intermediate value, in the middle panel, we choose $\alpha = 0.2$ for which the influence of rotation is already appreciable. The case shown in the top panel is for zero rotation and has already been considered in Paper I, we 
show it again here in order to highlight the effects of rotation by direct comparison. Increasing 
rotation (middle and bottom panels), we see a stabilizing effect on CDI, which progressively 
increases when $M_a$ decreases. This stabilizing effect of rotation has been already
discussed in Paper II \citep[see also][]{Carey09}. In this figure, the difference between the last two 
values of $\alpha$ may still be small, however, as we will see below, the behaviour can be noticeably
different at $M_a < 1$ for other values of $P_c$ and
$\gamma_c$, especially, in the limit $M_a\rightarrow 0$. By
contrast, the KH modes, occurring at larger values of $M_a$, are
essentially unaffected by rotation and, for this value of $P_c$,
are the dominant modes. Fig. \ref{fig:cdvar2} shows the same
kind of plots, but for a lower value of the pitch, $P_c=1$. In the top
panel (no rotation, $\alpha = 0$) we see that, as discussed in Paper
I, the CD mode increases its growth rate and moves towards larger
values of the wavenumber. Rotation has, as before, a stabilizing
effect, that becomes stronger as we decrease $M_a$. At zero rotation,
CDI is the dominant mode, its growth rate is independent from $M_a$ (for $M_a < 1$),
and is about an order of magnitude larger than the growth rate of
KHI. As we increase rotation, the
growth rate of CDI decreases and the stability boundary moves
towards smaller and smaller $k$ as $M_a$ is decreased. This decrease
in the level of CDI in Fig. \ref{fig:cdvar2} is most dramatic when $\alpha$ 
goes from zero to $0.2$ (as it is visible by comparing the top and 
middle panels), further increasing $\alpha$ to $1$ the decrease is then 
much less pronounced. As a result, for $\alpha = 1$, the KHI is
again the mode with the highest growth rate, which, however, has
changed only slightly relative to its value in the non-rotating    
case. The cases with the same high $\gamma_c=10$ and three different
values of the pitch, $P_c=10$, $P_c=1$ and $P_c=0.1$, are shown,
respectively, in Figs. \ref{fig:cdvar3}, \ref{fig:cdvar4} and
\ref{fig:cdvar5}. We note that the pitch measured in the jet rest
frame is given by $\gamma_c P_c$, so these three values would
correspond to $P_c=100$, $P_c=10$ and $P_c=1$ when measured in the
rest frame. (Additionally, we have to note that for $\gamma_c =
1.01$ there are no equilibrium solutions for $P_c=0.1$.) In the top
panel (no rotation) of Fig \ref{fig:cdvar3}, we therefore see that
the stability boundary of the CD modes moves further to the left,
i.e., towards smaller wavenumbers compared to the above case with
$\gamma_c=1.01$, because of the high value of the pitch measured in
the jet rest frame. Increasing rotation (middle and bottom panels),
below $M_a \sim 2$ the CD mode is stable (at least in the wavenumber
range considered). For lower values of the pitch in Figs.
\ref{fig:cdvar4} and \ref{fig:cdvar5}, in the absence of rotation, 
the stability limit of the CD modes shifts again to larger $k$ with
decreasing pitch. The effect of rotation is thus similar also in
this highly relativistic case as it is for $\gamma_c=1.01$, being
most remarkable when $\alpha$ increases from zero to 0.2. Notice
that, as discussed in Paper I, we have a splitting of the CD mode as
a function of wavenumber for $P_c=1$, this splitting is,
however, only present at zero rotation (top panel of Fig.
\ref{fig:cdvar4}). As noted above, the KH mode is essentially
unaffected by rotation. Therefore, except for the case with $P_c =
0.1$ and no rotation, the mode with the highest growth rate remains
the KHI. Finally, in Fig. \ref{fig:cdvar6} we show the case with
$P_c=0.01$, $\gamma_c=10$ and $\alpha=1$. For this value of $P_c$,
equilibrium is possible only at high values of $\gamma_c$ and the
allowed values of $\alpha$ cannot be much smaller than 1. The
behavior is similar to those discussed above, on one hand the CDI
tends to move towards higher wavenumbers and increase its growth
rate due to the decreasing value of $P_c$, on the other hand,
rotation has the usual stabilizing effect and, as a result, creates
an inclined stability boundary that moves towards smaller
wavenumbers as $M_a$ is decreased.

These results show that the effect of rotation is generally
stabilizing for the CD mode, at $M_a \lesssim 1$ (high
magnetization). An interesting question is then what happens in the
limit $M_a \rightarrow 0$. This limit, corresponding to the
force-free regime in relativistic jets, was investigated by
\citet{Pariev94,Pariev96,Lyubarski99} and \citet{Tomimatsu01}.
\citet{Tomimatsu01} derived the following condition for instability:
\begin{equation}\label{eq:tomimatsu}
|B_\varphi| > r \Omega_F B_z
\end{equation}
where $\Omega_F$ is the angular velocity of field lines that can be
expressed as (see Paper I)
\begin{equation}\label{eq:OmegaF}
\Omega_F  = \frac{v_\varphi}{r} - \frac{v_z}{P}.
\end{equation}
Using equation (\ref{eq:elec_field}) for the electric field $E_r$
and equation (\ref{eq:OmegaF}) for $\Omega_F$, the condition
(\ref{eq:tomimatsu}) can be written as
\begin{equation}\label{eq:tomimatsu2}
|B_\varphi| > |E_r|.
\end{equation}
The equilibrium condition (\ref{eq:radial_eq}) in the force-free
limit becomes
\begin{equation}
\frac{1}{2r}\frac{d(r^2H^2)}{dr} + \frac{r}{2}\frac{dB_z^2}{dr} = 0.
\end{equation}
If $B_z$ is constant, we have $H=0$ and $|B_\varphi| = |E_r|$.
According to Tomimatsu condition (\ref{eq:tomimatsu2}), in this
case, we are on the stability boundary and the system can be stable.
In fact, \citet{Pariev94, Pariev96} considered such a
situation and found stability. On the other hand, if $B_z$ decreases
radially outward, $H>0$, the Tomimatsu condition is satisfied and
there is instability. \citet{Lyubarski99} considered such
a case and indeed found instability, with a characteristic
wavenumber increasing inversely proportional to pitch. In our setup,
a constant $B_z$ corresponds to $\alpha = 1$, while a radially
decreasing vertical field corresponds to $\alpha<1$. From the above
figures we have seen that, in the presence of rotation, the
instability boundary moves towards smaller and smaller values of the
wavenumber as $M_a$ becomes low. However, it is hard to deduce from
this result how exactly the instability region changes along $k$
when approaching the force-free limit, $M_a\rightarrow 0$,
because of the limited interval of $M_a$ and $k$ values represented.
Nevertheless, we can see that, in general, at a given small $M_a\ll
1$, the stability boundary for $\alpha = 1$ tends to be at values of
$k$ smaller than those for $\alpha = 0.2$ (see e.g., Figs.
\ref{fig:cdvar2} and \ref{fig:cdvar5}), that is overall consistent
with the results of \citet{Pariev94, Pariev96,
Lyubarski99} and \citet{Tomimatsu01}.


\subsection{Centrifugal-buoyancy modes}

In Paper II, we demonstrated that in rotating non-relativistic jets,
apart from CD and KH modes, there exists yet another important class of
unstable modes that are similar to the Parker instability with the
driving role of external gravity replaced by the centrifugal force
\citep{Huang03} and analysed in detail their properties. At small
values of $k$, these modes operate by bending mostly toroidal field
lines, while at large $k$ they operate by bending poloidal field
lines. Accordingly, we labeled them the toroidal and poloidal
buoyancy modes \citep[see also][]{Kim00}. In this subsection, we
investigate how the growth of these modes is affected by
relativistic effects.

\subsubsection{Toroidal buoyancy mode}

\begin{figure}
\centering
\includegraphics[width=\columnwidth]{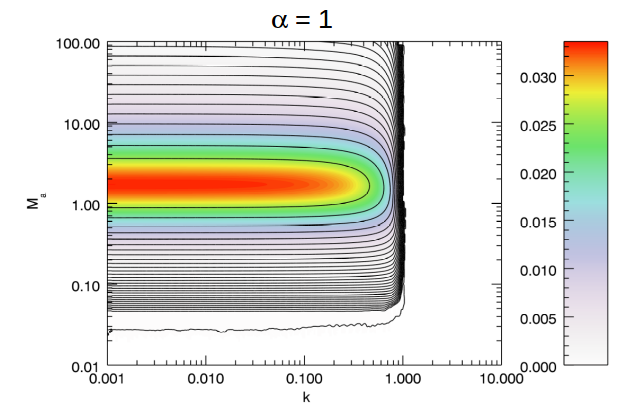} 
\caption{\small Distribution of the growth rate, $-{\rm Im}(\omega)$, of the toroidal
buoyancy mode as a function of the wavenumber $k$ and $M_a$ for
$\gamma_c = 1.01$, $\alpha = 1$ and $P_c = 1$.} \label{fig:tor1}
\end{figure}

The toroidal buoyancy mode operates at small values of $k$ and, in
fact, its instability is present only for wavenumbers $k<k_c$, where
the high wavenumber cutoff $k_c$ depends on the pitch parameter and
satisfies the condition $k_c P_c \sim 1$. Fig. \ref{fig:tor1}
presents the typical behaviour of the growth rate of the toroidal
mode as a function of $k$ and $M_a$ for $P_c=1$, $\gamma_c =1.01$
and $\alpha = 1$. It is seen that in the unstable region, the growth
rate is essentially independent from $k$ and reaches a maximum for
$M_a$ slightly larger than 1. At smaller and larger values of $M_a$
the mode is stable or has a very small growth rate, depending on the
parameters, as we will see below.

\begin{figure}
\centering
\includegraphics[width=\columnwidth]{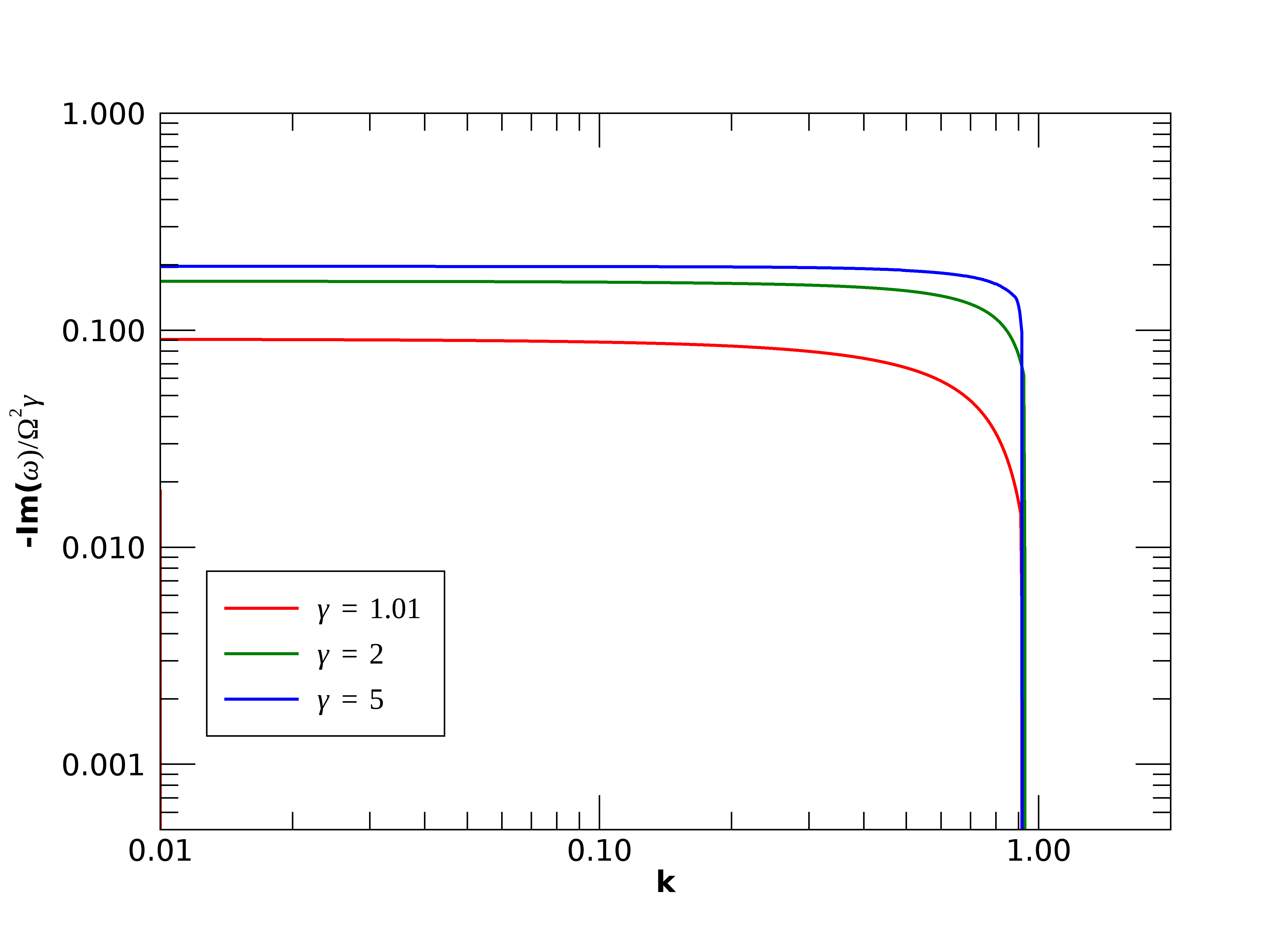} 
\caption{\small Growth rate of the toroidal mode as a function of
$k$ for $\alpha=1$, $P_c=1$ and $\gamma_c=1.01,2,5$. The growth rate
is normalized by $\gamma_c \Omega_c^2$. The three different curves
refer to three different values of $\gamma_c$ as indicated in the
legend. $M_a$ is also different for the three curves and for each
curve it has the value at which the maximum growth rate is reached
at a fixed $k$: $M_a = 1.5,~4.5,~15$, respectively,
for $\gamma_c=1.01,~2,~5$.} \label{fig:tor2}
\end{figure}

\begin{figure*}
\centering
\includegraphics[width=7.5cm]{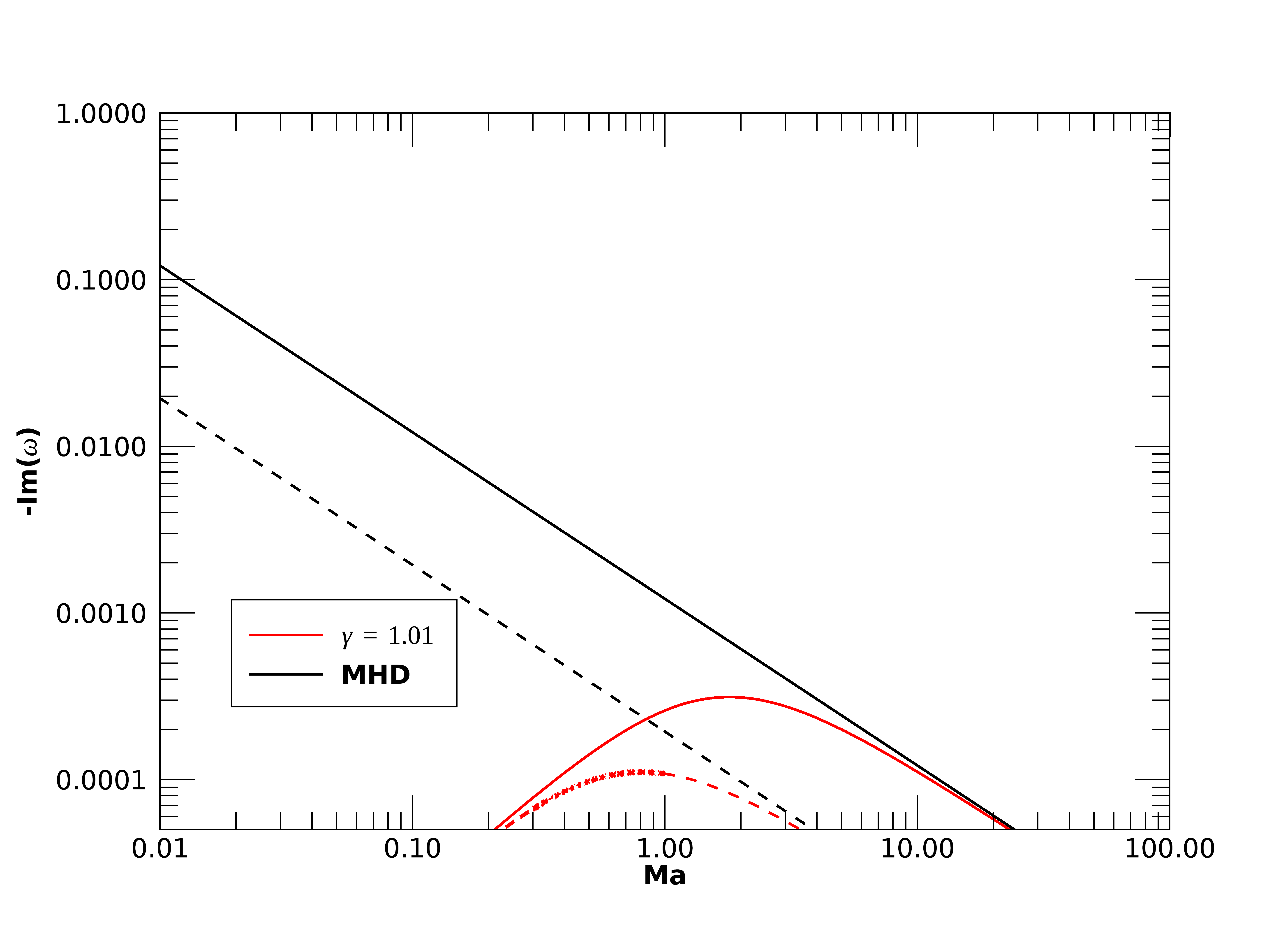} 
\includegraphics[width=7.5cm]{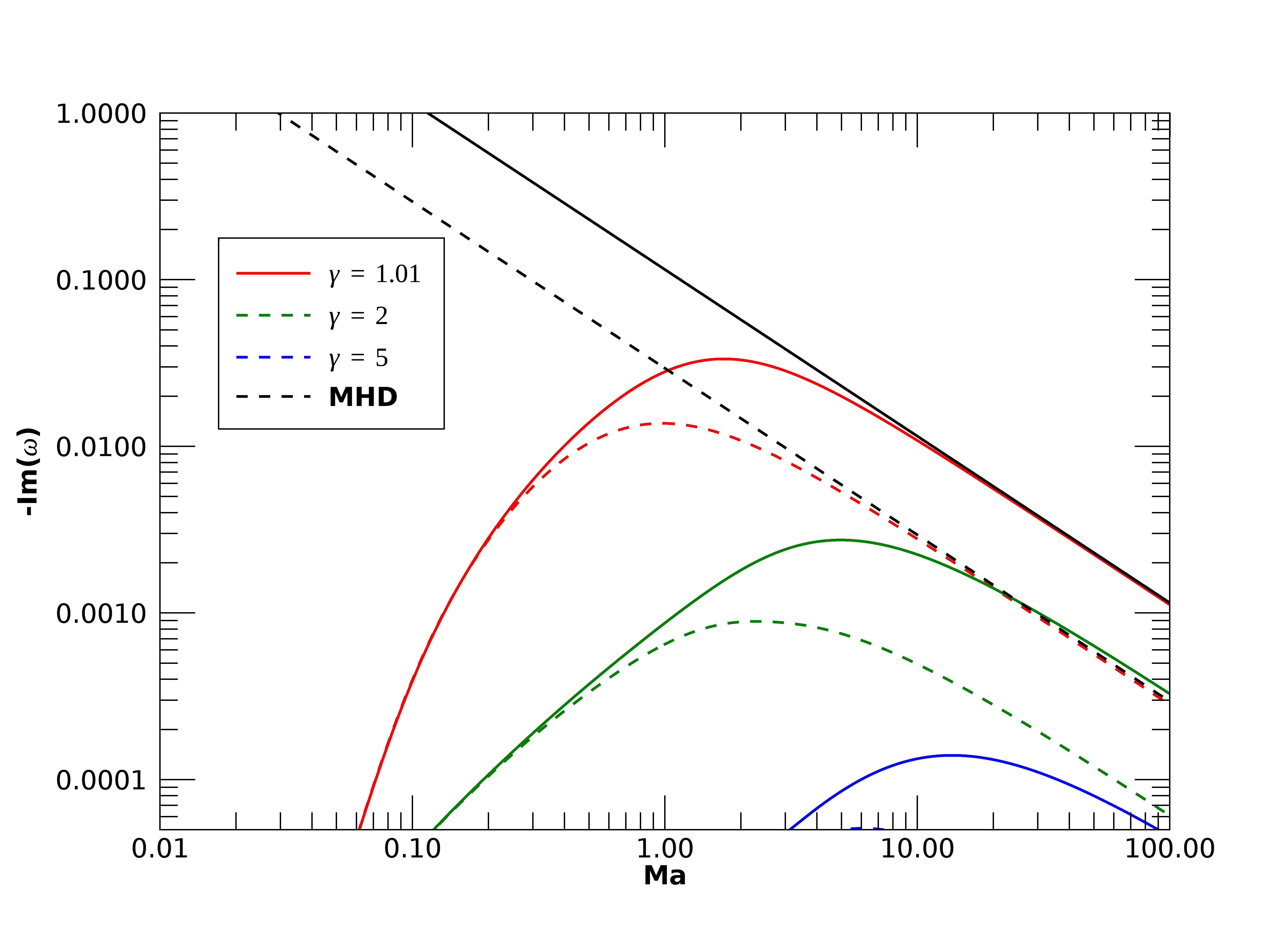} 
\includegraphics[width=7.5cm]{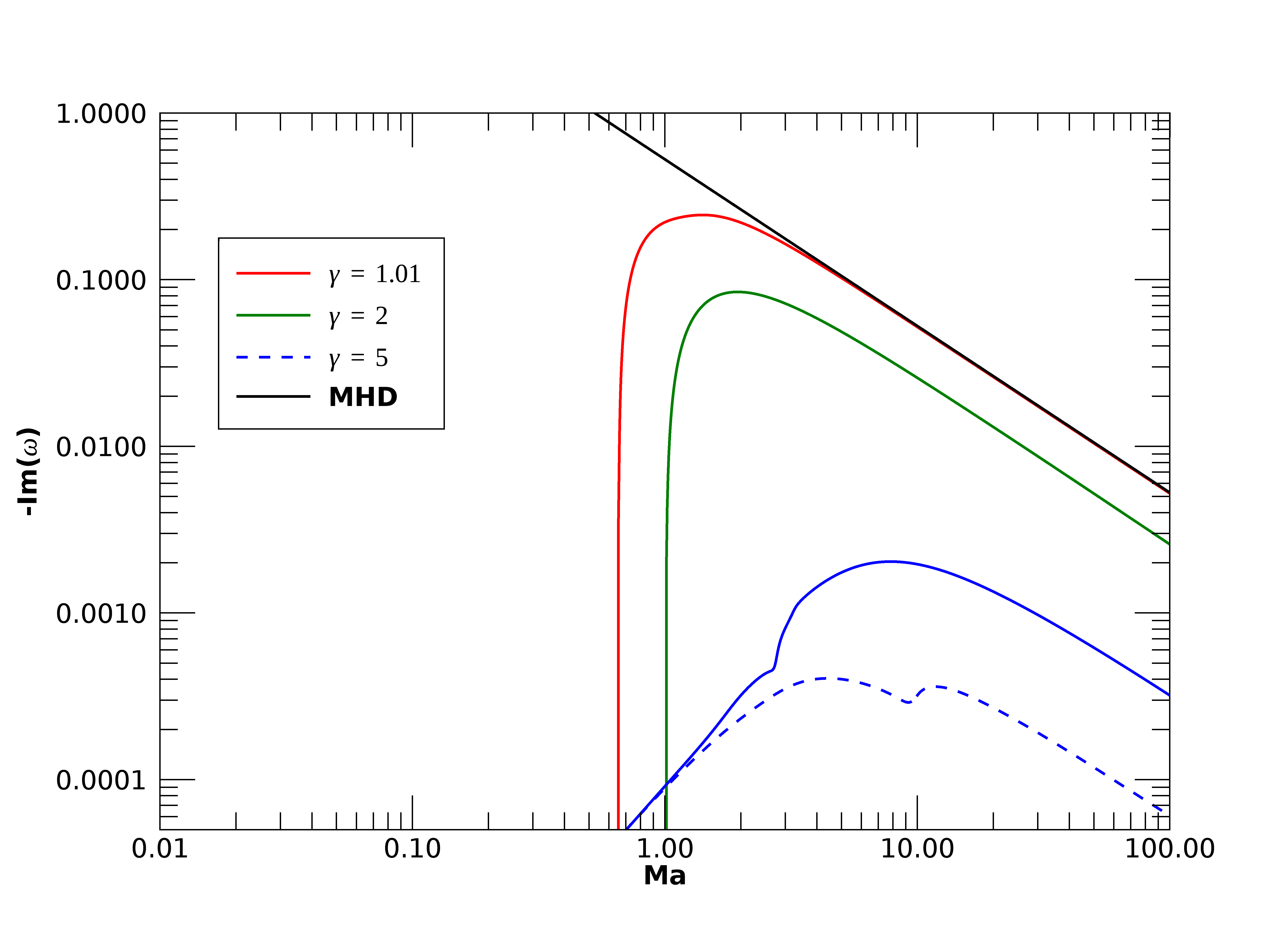} 
\includegraphics[width=7.5cm]{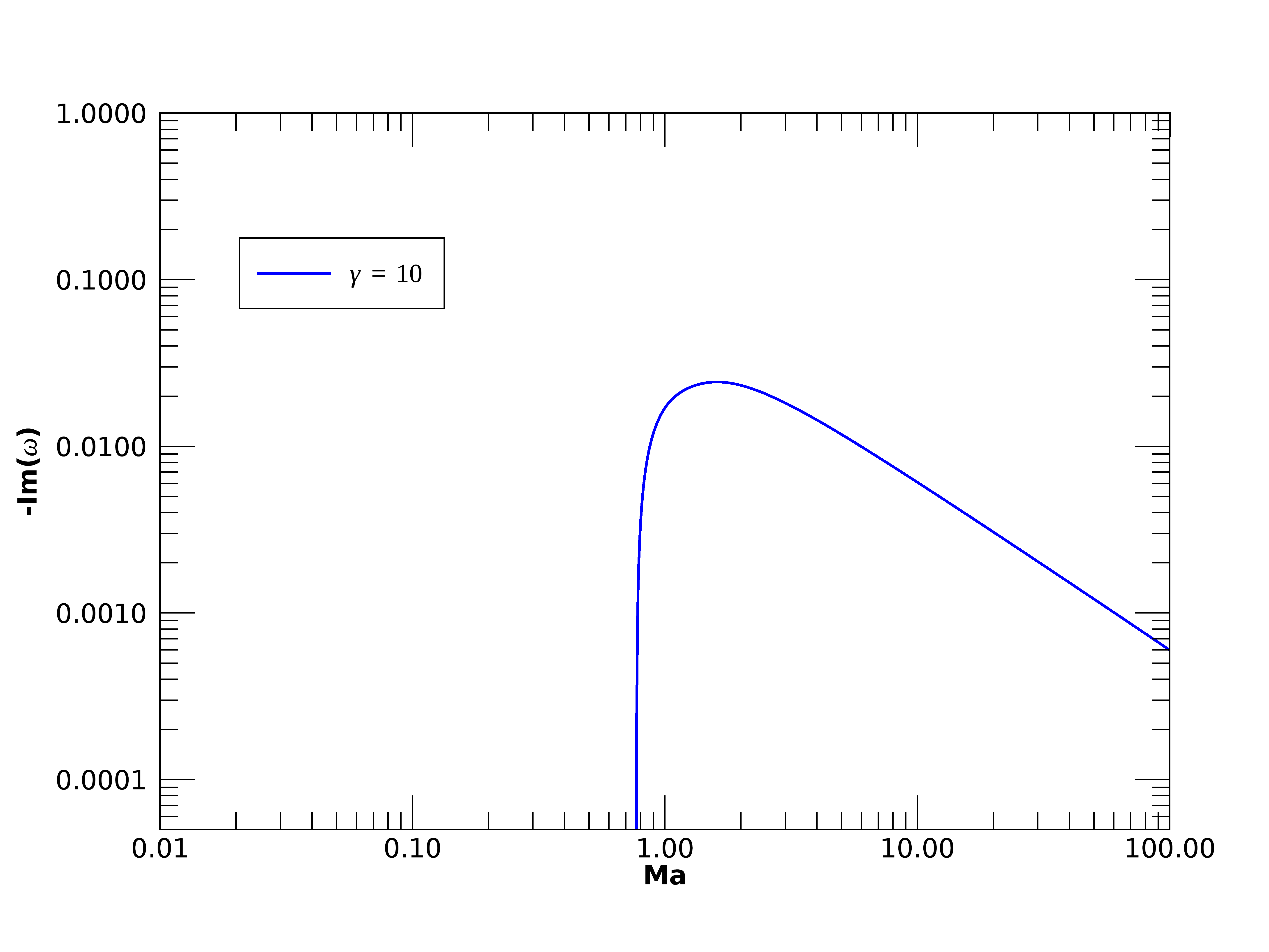} 
\caption{\small  Growth rate of the toroidal mode as a function of
$M_a$. The four panels refer to different values of the pitch
parameter $P_c$ (Top left: $P_c = 10$;  Top right: $P_c = 1$; Bottom
left: $P_c = 0.1$; Bottom right: $P_c = 0.01$). Solid curves are for
$\alpha = 1$, dashed curves are for $\alpha=0.2$, curves with
different colours refer to different values of $\gamma_c=1.01,2,5,10$
as indicated in the legend of each panel. The black lines represent 
the growth rate calculated with non-relativistic ideal MHD equations 
at zero thermal pressure and the same pitch and $\alpha$ in the respective panel.  If the dashed curve is
absent in any panel, this means that the equilibrium with
$\alpha=0.2$ is not possible for that pair of $P_c$ and $\gamma_c$
associated with this panel. The wavenumber is $k=0.01$ in all the cases,
although the growth rate is essentially independent of this
value.}\label{fig:tor3}
\end{figure*}

In Paper II, we showed that the growth rate of the centrifugal-buoyancy 
modes scales approximately as $\Omega_c^2$. In the present
case, we have to take into account the relativistic effects and, as
the mode tends to be concentrated inside the jet, we have to
consider quantities measured in the rest frame of the jet. The
growth rate in the rest frame should scale as in the
non-relativistic case, i.e., as the square of the rotation frequency
in this frame. Since the growth rate in the jet rest frame is ${\rm
Im}(\omega') = \gamma_c {\rm Im}(\omega)$, while the rotation
frequency is $\Omega'_c = \gamma_c \Omega_c$, we can write the
scaling law in the lab frame
\begin{equation}\label{eq:scaling}
-{\rm Im}(\omega) \sim \gamma_c \Omega_c^2,
\end{equation}
In Fig. \ref{fig:tor2}, we plot the growth rate normalized according
to this scaling, $-{\rm Im}(\omega)/(\gamma_c\Omega_c^2)$, as a
function of the wavenumber for $\alpha = 1$, $P_c = 1$ and three
different values of $\gamma_c=1.01,2,5$. For each $\gamma_c$, we
choose the value of $M_a$, which corresponds to the maximum growth
rate at a given $k$, these values are reported in the figure
caption. For each curve, the value of $\Omega_c$ is also different
and equal to $0.6$ for $\gamma_c=1.01$, $0.09$ for $\gamma_c = 2$
and $0.01$ for $\gamma_c = 5$. The growth rate in the unstable range
is independent from the wavenumber, as seen in Figs. \ref{fig:tor1}
and \ref{fig:tor2}, and the scaling law (\ref{eq:scaling})
reproduces quite well the behaviour of the growth rate at different
$\gamma_c$: the corresponding curves come close to each other 
(collapse) when the normalized growth rate is plotted.

To study in more detail the dependence of the toroidal buoyancy
instability on the jet flow parameters, in Fig. \ref{fig:tor3}, we
plot the growth rate as a function of $M_a$ for a given $k$ and
various $P_c$ and $\gamma_c$. Although the value of $k$ is fixed in
 these plots, we recall that, as we have seen above, there is no
dependence of the growth rate on $k$ when  $k$ is sufficiently
smaller than the cut-off value. So, the curves in this figure would not
change for other choices of the unstable wavenumber. The four
different panels correspond to different values of the pitch
parameter (top left: $P_c=10$, top right: $P_c = 1$, bottom left:
$P_c = 0.1$, bottom right: $P_c = 0.01$). In each panel, different
colours refer to different values of $\gamma_c=1.01,2,5,10$, while
the solid curves are for $\alpha = 1$ and dashed ones for $\alpha =
0.2$. Not all curves are present in all panels because, as discussed
in subsection 2.1, there are combinations of parameters for which
equilibrium is not possible.   At large values of $M_a \gtrsim 10$ (low
magnetization), the growth rate decreases as $1/M_a$ for all values of 
$\gamma_c$, $P_c$ and $\alpha$ given in these panels. In particular, at $\gamma_c = 1.01$ the 
behaviour coincides with the non-relativistic MHD case (black lines in Fig. \ref{fig:tor3} calculated with ideal non-relativistic MHD equations at zero thermal pressure, as in Paper II). 
This behaviour can be explained as follows. At large 
$M_a$, rotation is balanced by magnetic forces and hence both decrease with increasing $M_a$. 
As a result, the growth rate of the centrifugal-buoyancy modes decreases too, because it scales 
with the square of the rotation frequency (Eq. \ref{eq:scaling}).
As $M_a$ decreases, the growth rate first increases more 
slowly, reaches a maximum and then decreases at small $M_a\ll 1$, as we
approach the force-free limit where the centrifugal instabilities should   
eventually disappear. The behaviour of the growth rate as a function
of $\alpha$, $P_c$ and $\gamma_c$ can be understood from the same scaling
with the rotation frequency discussed above. Increasing $\alpha$,
the rotation frequency increases as well and, consequently, the growth
rate. A decrease in the pitch also leads to an increase of the
rotation frequency and, therefore, of the growth rate. Finally,
understanding the dependence on $\gamma_c$ is more complex since we
have to take into account the transformation of all the quantities
from the lab frame to the jet frame. The pitch in the jet frame is
given by $\gamma_c P_c$ and is, therefore, larger for larger
$\gamma_c$ and the rotation rate is consequently smaller. In
addition, the growth rate measured in the lab frame is smaller by a
factor of $\gamma_c$ than the growth rate measured in the jet frame,
as a result, the growth rate strongly decreases with $\gamma_c$.

\begin{figure}
\centering
\includegraphics[width=\columnwidth]{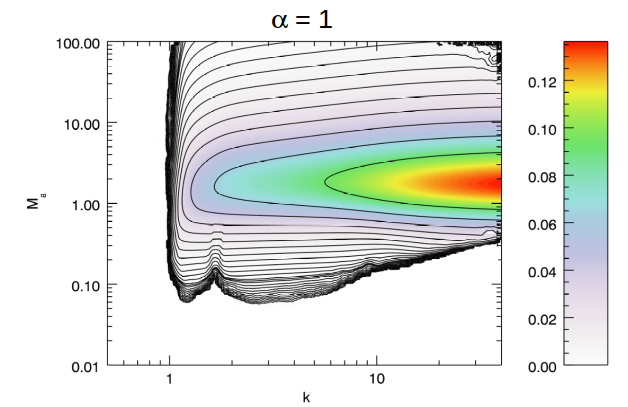} 
\caption{\small Distribution of the growth rate, $-{\rm Im}(\omega)$, of the poloidal
buoyancy mode as a function of the wavenumber $k$ and $M_a$ for
$\gamma_c = 1.01$, $\alpha = 1$ and $P_c = 1$.}\label{fig:pol1}
\end{figure}

\begin{figure}
\centering
\includegraphics[width=\columnwidth]{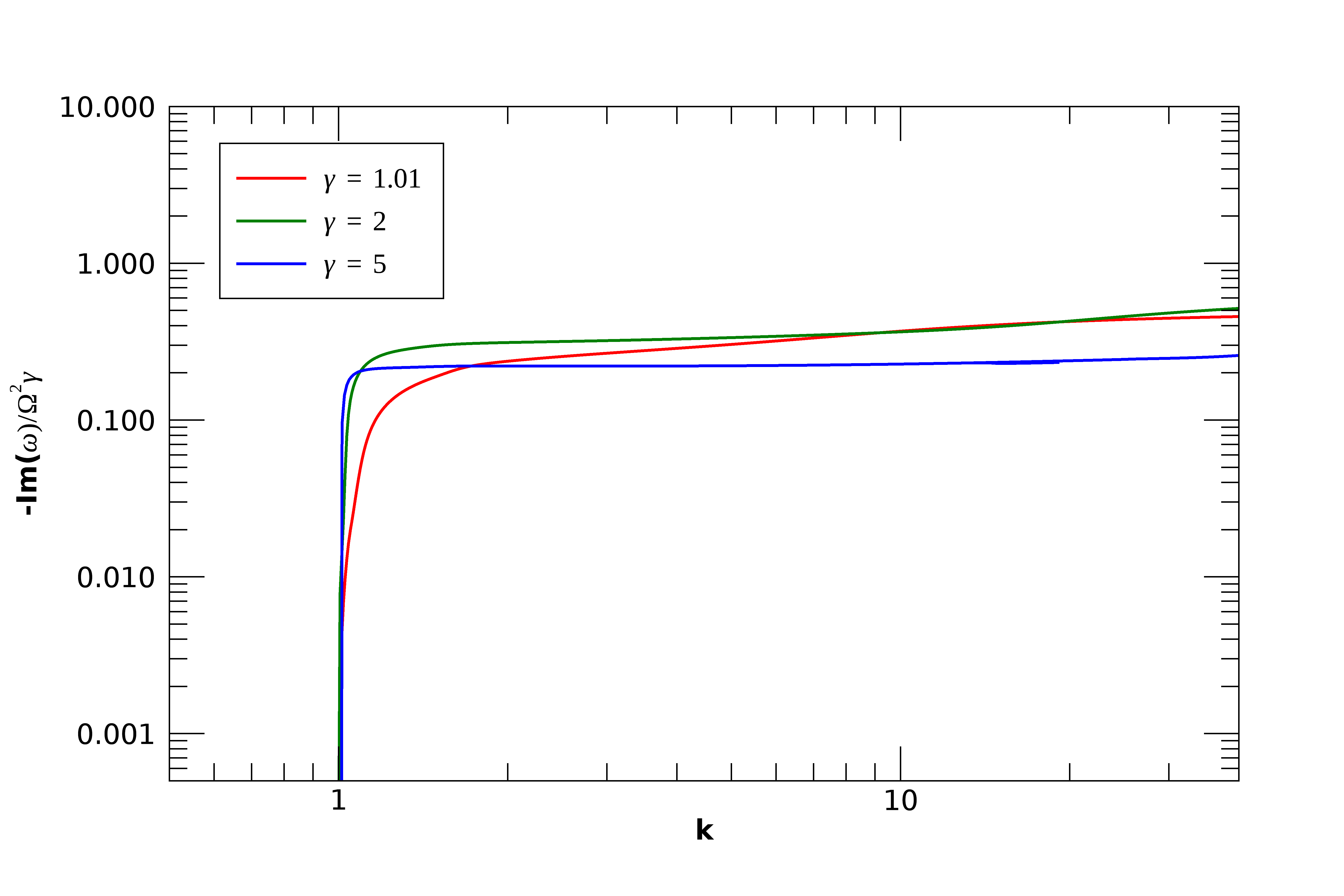} 
\caption{\small Growth rate of the poloidal buoyancy mode as a
function of $k$ for $\alpha=1$, $P=1$ and $\gamma_c=1.01,2,5$. The
growth rate is normalized by $\gamma_c\Omega_c^2$. The three
different curves refer to three different values of $\gamma_c$ as
indicated in the legend. $M_a$ is different for the three curves and
for each curve has the value at which the maximum growth rate is
found at fixed $k$: $M_a =1.8,~5.72,~14.2$,
respectively, for $\gamma_c=1.01,~2,~5$.}
\label{fig:pol2}
\end{figure}

\begin{figure*}
\centering
\includegraphics[width=7.5cm]{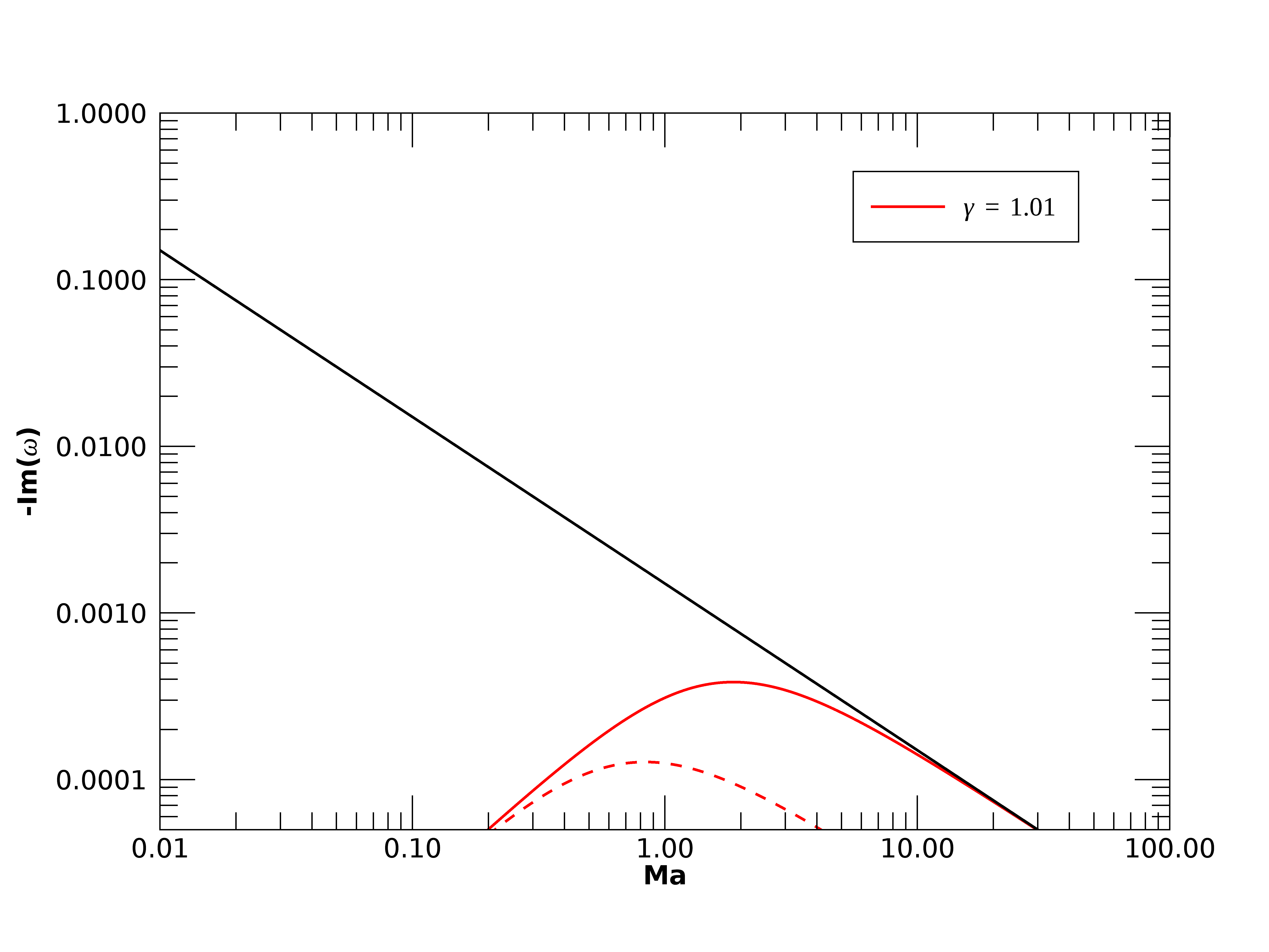} 
\includegraphics[width=7.5cm]{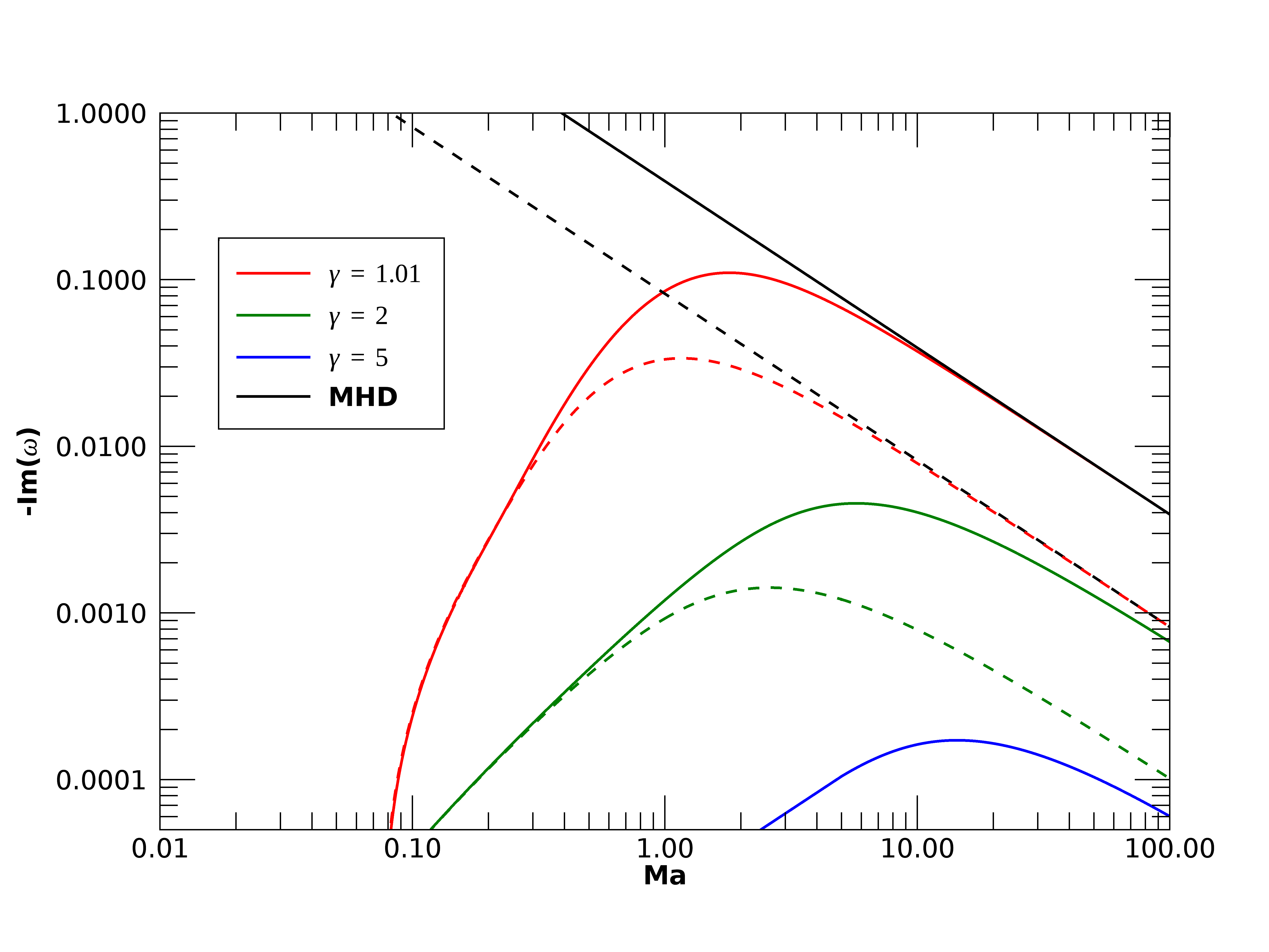} 
\includegraphics[width=7.5cm]{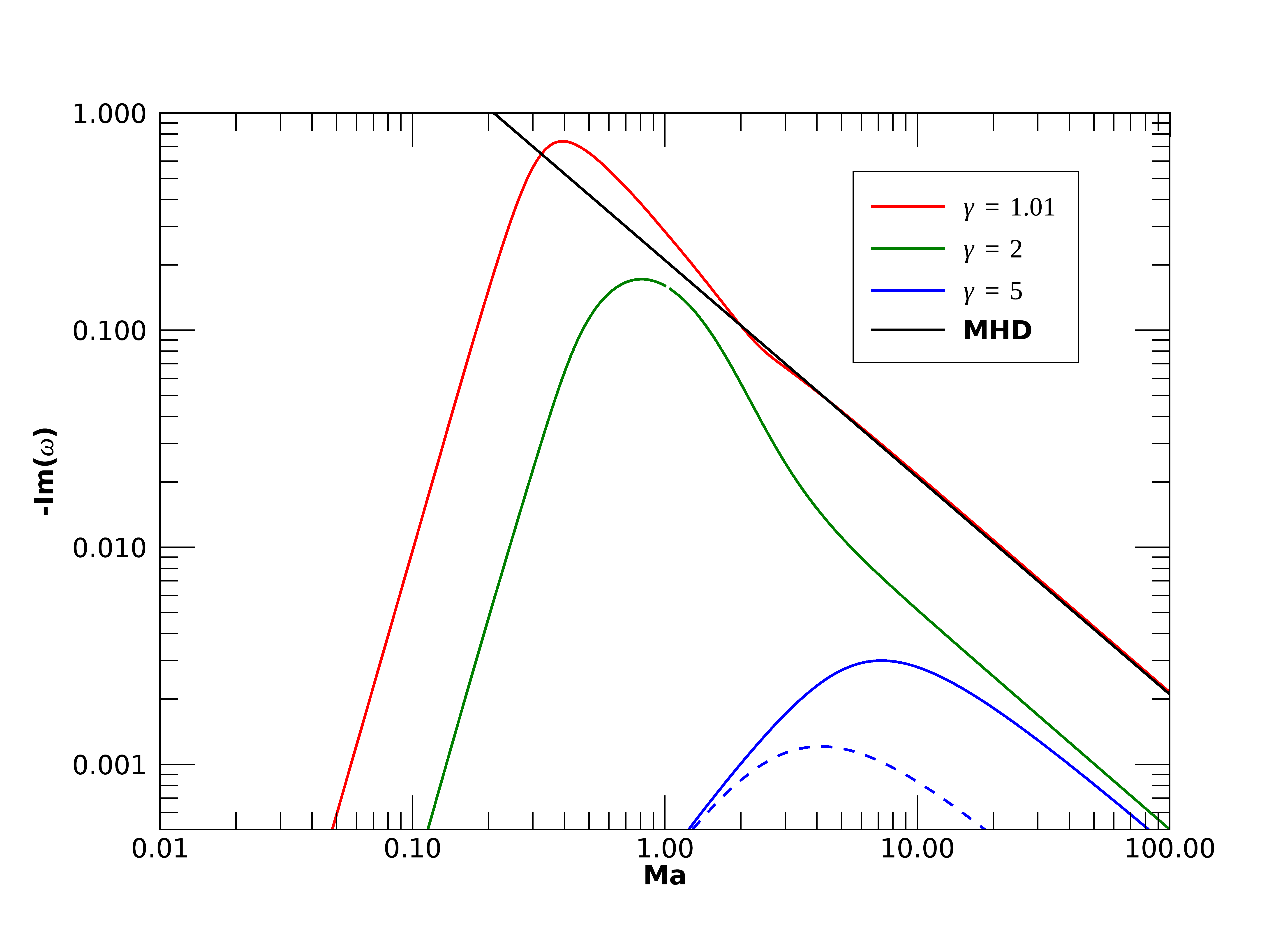} 
\includegraphics[width=7.5cm]{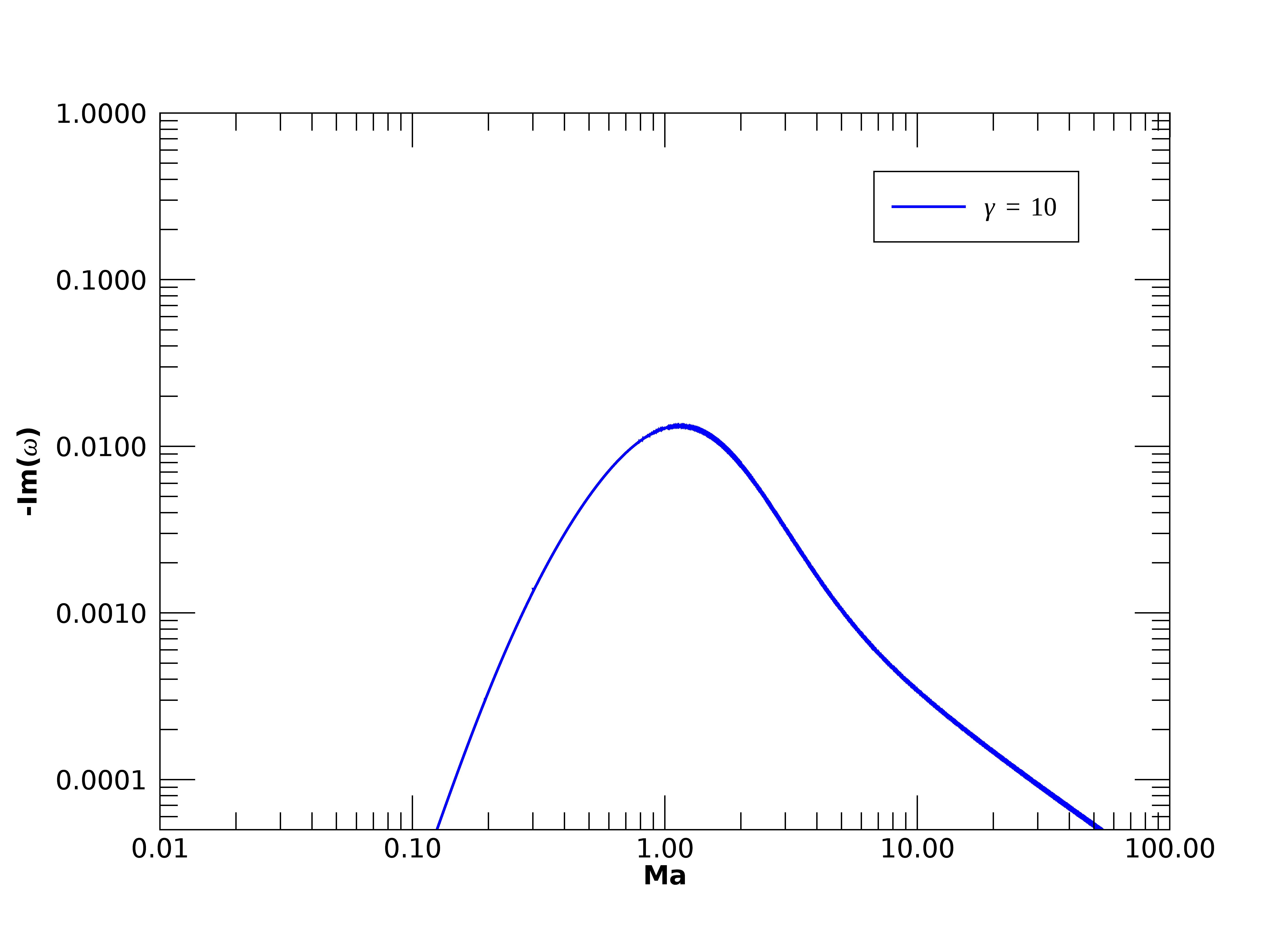} 
\caption{\small Growth rate of the poloidal buoyancy mode as a
function of $M_a$. The four panels refer to different values of the
pitch parameter $P_c$ (Top left: $P_c = 10$;  Top right: $P_c = 1$;
Bottom left: $P_c = 0.1$; Bottom right: $P_c = 0.03$). Solid curves
are for $\alpha = 1$, dashed curves are for $\alpha=0.2$, curves
with different colours refer to different values of
$\gamma_c=1.01,2,5,10$ as indicated in the legend.   As in Fig. \ref{fig:tor3}, 
the black lines represent the results obtained using ideal non-relativistic 
MHD equations for the same pitch and $\alpha$ in each panel.  The dashed curves are absent 
in those panels with such a pair of $P_c$ and $\gamma_c$ that do not allow the equilibrium with 
$\alpha=0.2$. For computation reasons, the wavenumber $k$ is different for each panel (Top left:
$k=5$; Top right: $k=10$; Bottom left: $k=14$; Bottom right: $k=40$), but corresponds to the 
regime where the growth rate practically no longer depends on it.}\label{fig:pol3}
\end{figure*}

\subsubsection{Poloidal buoyancy mode}

Another type of the centrifugal-buoyancy mode existing in the
rotating jet is the poloidal buoyancy mode. As mentioned above, this
mode operates at large $k$ by bending mostly poloidal field lines.
In Fig. \ref{fig:pol1}, we present the behaviour of its growth rate
as a function of the wavenumber $k$ and $M_a$ for $P_c=1$, $\gamma_c
=1.01$ and $\alpha = 1$. This figure is almost specular with respect
to Fig. \ref{fig:tor1} and shows that the poloidal buoyancy
instability first starts from the same cutoff wavenumber $k_cP_c
\sim 1$ and extends instead to larger wavenumbers, $k>k_c$, having 
the growth rate somewhat larger than that of the toroidal buoyancy
mode. At fixed $k$, it is concentrated in a certain range of $M_a$,
with the maximum growth rate being achieved around $M_a\sim 1$ at
every $k$ and decreasing at large and small $M_a$. At a given finite
$M_a$, in the unstable region, the growth rate initially increases
with $k$ and then tends to a constant value at $k\gg k_c$, as also 
seen in Fig. \ref{fig:pol2}. Like the toroidal buoyancy mode, the
poloidal buoyancy mode obeys the same scaling law
(\ref{eq:scaling}), because it is also determined by the centrifugal
force. This is confirmed by Fig. \ref{fig:pol2}, which presents the
growth rate as a function $k$ for $\alpha = 1$, $P_c = 1$ and three
different values $\gamma_c=1.01,2,5$, while $M_a$ is chosen for each
curve such that to yield the maximum the growth rate at a given wavenumber.
Thus, modification (reduction) of the growth of both centrifugal-buoyancy
instabilities in the relativistic case compared to the
non-relativistic one is in fact mainly due to the time-dilation 
effect -- in the jet rest frame their growth rate is determined by
$\Omega_c'^2$ in this frame, as it is in the non-relativistic case.

Fig. \ref{fig:pol3} shows the behaviour of the poloidal buoyancy mode
as a function of $M_a$ for fixed, sufficiently large values of $k$, when the
instability is practically independent of it (see Fig.
\ref{fig:pol2}). The values of $\alpha$, $\gamma_c$ and $P_c$ are the same as
used in Fig. \ref{fig:tor3} except that in the bottom right panel we used
$P_c = 0.03$ instead of 0.01 since for $P_c$ slightly below 0.03 the
poloidal mode becomes stable, when the cutoff wavenumber, $k_c$,
which is set by the pitch, becomes larger than the fixed wavenumber
($k=40$) used in this panel. Overall, the dependence of the growth
rate of the poloidal buoyancy mode on these parameters is quite
similar to that of the toroidal one described above. In particular, at large $M_a$ the growth rate 
varies again as $1/M_a$ at all other parameters, coinciding at $\gamma_c=1.01$ with the behaviour 
in the non-relativistic case (black lines). It then increases with decreasing $M_a$, reaches a
maximum and decreases at small $M_a$. The lower is $\gamma_c$ the
higher is this maximum and the smaller is the corresponding $M_a$.
On the other hand, with respect to pitch, the highest growth is
achieved at $P_c\sim 0.1$ at all values of $\gamma_c$ considered. Due
to the above scaling with the jet rotation frequency, the growth
rate also increases with $\alpha$. This behaviour of the poloidal
mode instability as a function of $\alpha$, $P_c$ and $\gamma_c$  can 
be explained by invoking similar arguments as for the toroidal mode in the previous subsection.

\section{Summary}
\label{summary}

We have investigated the stability properties of a relativistic
magnetized rotating cylindrical flow, extending the results obtained
in Papers I and II. In Paper I, we neglected rotation, while in
Paper II we did not consider the presence of the longitudinal flow
and relativistic effects, here we considered the full case, still
remaining however in the limit of zero thermal pressure. In the
first two papers, we discussed several modes of instabilities that
in the present situation all exist. The longitudinal flow velocity
gives rise to the KHI, the toroidal component of the magnetic field
leads to CDI, while the combination of rotation and magnetic fields 
gives rise to unstable toroidal and poloidal centrifugal-buoyancy modes. The instability behaviour depends, of course, on the chosen equilibrium configuration and our results can be considered representative of an equilibrium
configuration characterized by a distribution of current
concentrated in the jet, with the return current assumed to be
mainly found at very large distances. Not all combinations of
parameters are allowed: there are combinations for which equilibrium
solution does not exist. More precisely, for any given rotation
rate, there is a minimum value of the pitch, below which no
equilibrium is possible. Increasing the rotation rate, this minimum
value of the pitch decreases.

The behaviour of KHI and CDI is similar to that discussed in Paper I,
at high values of $M_a$ we find the KHI, while at low values we find
the CDI. The KHI is largely unaffected by rotation, which, on the
contrary, has a strong stabilizing effect on the CDI. Decreasing
$M_a$ and increasing rotation, the unstable region (stability
boundary) progressively moves towards smaller axial wavenumbers. A
decreasing value of the pitch, on the other hand, moves the
stability limits towards larger axial wavenumbers. For relativistic
flows, we have also to take into account that the pitch measured in
the jet rest frame, which determines the behaviour of the CDI, is
$\gamma_c$ times the value measured in the laboratory frame, therefore
relativistic flows, with the same pitch, are more stable. 
 In Paper I, we found a scaling law showing that the growth rate 
of CDI strongly decreases with
$\gamma_c$, while an even stronger stabilization
effect is found here due to the combination of relativistic effects and rotation.
Extrapolating the behaviour found at low values of $M_a$ to the limit
$M_a \rightarrow 0$, the results are in agreement with Tomimatsu
condition \citep{Tomimatsu01}, applicable to the force-free limit.

Rotation drives centrifugal-buoyancy modes: the toroidal buoyancy
mode at low wavenumbers and the poloidal buoyancy mode at high axial
wavenumbers. Apart from the different range of these wavenumbers,
they have a similar behaviour and their growth rate scales with the
square of the rotation frequency, which, in turn, increases as the
pitch decreases; therefore they become important at low values of
the pitch. In the unstable range, the growth rate is independent
from the wavenumber. As the magnetic field increases they tend to be
more stable. The same happens increasing the Lorentz factor of the
flow.

In this paper, we have considered only the $m=1$ mode, as this mode
is thought to be the most dangerous one for jets. Higher order
modes $(m>1)$ can trigger instabilities internally in the jet,
instead of a global kink. These perturbations can cause inherent
breakup of current sheets, reconnection, etc. So, these modes are
also interesting to study in the future. If the global jet can remain stable
for long time scales and large distances, locally, higher order
modes can cause local instabilities that can or cannot disrupt the
jet.

In summary, rotation has a stabilizing effect on the CDI which
becomes more and more efficient as the magnetization is increased.
Rotation on the other hand drives centrifugal modes, which, however,
are also stabilized at high magnetizations. Finally, relativistic
jet flows tend to be more stable compared to their non-relativistic
counterparts.

\section*{Acknowledgments}
This project has received funding from the European Union's Horizon
2020 research and innovation programme under the Marie Sk{\l}odowska
-- Curie Grant Agreement No. 795158 and from the Shota Rustaveli
National Science Foundation of Georgia (SRNSFG, grant number
FR17-107). GB, PR, AM acknowledge support from PRIN MIUR 2015 (grant
number 2015L5EE2Y) and GM from the Alexander von Humboldt Foundation
(Germany).

\onecolumn
\appendix
\section{Equations of the eigenvalue problem for the jet stability}
\label{appendix}

Considering small perturbations of the velocity, electric and
magnetic fields, $\vec{v}_1$, $\vec{E}_1$, $\vec{B}_1 \propto
\exp({\rm i}\omega t-{\rm i}m\varphi-{\rm i}kz)$, about the
equilibrium state with $\vec{v}=(0,v_{\varphi},v_z)$,
$\vec{E}=(E_r,0,0)$, $\vec{B}=(0,B_{\varphi},B_z)$ described in
subsection 2.1, and linearizing the main equations
(\ref{eq:drho/dt})-(\ref{eq:dE/dt}), after some algebra, we arrive
at the following system of linear differential equations for the
radial displacement $\xi_{1r}=-{\rm i}v_{1r}/\tilde{\omega}$ and
electromagnetic pressure $\Pi_1=\vec{B}\cdot
\vec{B}_1-\vec{E}\cdot\vec{E}_1=B_{\varphi}B_{1\varphi}+B_zB_{1z}-E_rE_{1r}$
(see Appendix A of Paper I for the detailed derivations),
\begin{equation} \label{eq:dxi/dr}
\left. D\frac{d\xi_{1r}}{dr}=\left(C_1+\frac{C_2-D
k_B'}{k_B}-\frac{D}{r}\right)\xi_{1r} - C_3 \Pi_1 \right.
\end{equation}
\begin{equation} \label{eq:dPhii/dr}
D\frac{d\Pi_1}{dr} = \left[A_1
D-\frac{\rho\gamma^2v_{\varphi}^2}{r}\left(C_1+\frac{C_2-D
k'_B}{k_B}\right) + \frac{C_4}{r} + C_5\right]\xi_{1r}+
\left[\frac{1}{r}\left(\rho\gamma^2v_{\varphi}^2C_3-2D + C_6\right)
+ C_7 \right]\Pi_1,
\end{equation}
where
\[
D\equiv\left(\frac{B^2}{\rho\gamma^2}+1\right)B^2\tilde{\omega}^2+\frac{k_BB^2}{\rho\gamma^2}\left[2\tilde{\omega}(\vec
{v}\cdot\vec{B})-\frac{k_B}{\gamma^2}\right],
\]
\[
\tilde{\omega} \equiv \omega-\frac{m}{r}v_{\varphi}-kv_{z},\qquad
k_B\equiv \frac{m}{r}B_{\varphi}+kB_z, \qquad k_B'\equiv
\frac{dk_B}{dr}, \qquad B^2=B_{\varphi}^2+B_z^2.
\]
The other quantities, $A_1, C_1, C_2, C_3, C_4, C_5, C_6, C_7$
contained in these equations, also depend on the chosen profile of
the equilibrium solution, and their explicit forms are derived in
Appendix B of Paper I. They are rather long expressions and we do
not give them here.

The boundary conditions near the jet axis and at large radii are
also derived in Appendix C and D of Paper I and are as follows. In
the vicinity of the jet axis, $r\rightarrow 0$, the regular
solutions have the form $\xi_{1r}\propto r^{|m|-1}, \Pi_1\propto
r^{|m|}$ $(|m|\geq 1)$ and their ratio is
\begin{equation}\label{eq:bound_in}
\frac{\Pi_1}{\xi_{1r}}=\frac{r\rho\gamma_c^2}{mB_c^2}\left[{\rm
sign}(m)D+2B_{z}^2\left(\tilde{\omega}v'_{\varphi}+\frac{B^2}{\rho\gamma^2}\left(\omega-m\Omega_F\right)\Omega_F+\frac{B^2B'_{\varphi}k_B}{\rho\gamma^2B_{z}^2}\right)\right]_{|r=0}.
\end{equation}
Here $\Omega_F$ is the angular velocity of magnetic field lines
given by equation (\ref{eq:OmegaF}).

At large radii, $r\rightarrow \infty$, the solution represents
radially propagating waves that vanish at infinity. The
electromagnetic pressure perturbation of these waves is given by the
Hankel function of the first kind, $\Pi_1=H^{(1)}_{\nu}(\chi r)$,
where
\[
\chi^2=\frac{\rho+B_{z}^2}{B_{z}^2}\omega^2-k^2,~~~~\nu^2=
m^2+\frac{\rho B_{\varphi c}^2\omega^2}{B_{z}^4}, 
\]
with the leading term of the asymptotic expansion at $r\rightarrow
\infty$
\begin{equation}\label{eq:bound_out}
\Pi_1=H_{\nu}^{(1)}(\chi r)\simeq \sqrt{\frac{2}{\pi \chi
r}}\exp{\left[{\rm i}\left(\chi r-\frac{\nu
\pi}{2}-\frac{\pi}{4}\right)\right ]}.  
\end{equation}
In this expression, the complex parameter $\chi$ can have either positive or negative
sign,
\begin{equation}\label{eq:p1_asympt}
\chi=\pm\sqrt{\frac{\rho+B_{z}^2}{B_{z}^2}\omega^2-k^2}.
\end{equation}
Requiring that the perturbations decay at large radii, in equation (\ref{eq:p1_asympt}), we choose the root that has a
positive imaginary part, ${\rm Im}(\chi)>0$. These perturbations are
produced within the jet and hence at large radii should have the
character of radially outgoing waves. This implies that the real
parts of $\chi$ and $\omega$ should have opposite signs, ${\rm
Re}(\omega){\rm Re}(\chi)<0$ (Sommerfeld condition), in order to
give the phase velocity directed outwards from the jet. As a result,
the asymptotic behaviour of the displacement $\xi_{1r}$ can be
readily obtained from $\Pi_1$ correct to $O(r^{-3})$
\[
\xi_{1r}=\frac{\Pi_1}{\omega^2(\rho+B^2)-k_B^2}\left({\rm
i}\chi-\frac{1}{2r}\right). 
\]
The main equations (\ref{eq:dxi/dr}) and (\ref{eq:dPhii/dr})
together with the above boundary conditions at small (\ref{eq:bound_in}) and large (\ref{eq:bound_out}) radii are solved via shooting method to find the eigenvalues of $\omega$.

\label{lastpage}

\end{document}